\documentstyle[12pt]{article}

\begin{document}

\begin{flushright}
gr-qc/0205125 \\
May 2002
\end{flushright}

\begin{centering}
{\leftskip=2in \rightskip=2in
{\large \bf On the fate of Lorentz symmetry in}}\\
{\leftskip=2in \rightskip=2in
{\large \bf loop quantum gravity and noncommutative spacetimes}}\\
\bigskip
\bigskip
\medskip
{\bf Giovanni AMELINO-CAMELIA}\\
\bigskip
{\it Dipartimento di Fisica, Universit\'{a} ``La Sapienza", P.le Moro 2,
I-00185 Roma, Italy}\\
{\it Perimeter Institute for Theoretical Physics,
Waterloo, Canada N2J 2W9}

\end{centering}

\vspace{0.7cm}
\begin{center}
{\bf ABSTRACT}
\end{center}

\baselineskip 11pt plus .5pt minus .5pt

{\leftskip=0.6in \rightskip=0.6in
{\footnotesize Motivated by the remarkable sensitivity levels of
the Lorentz-symmetry tests
at some presently-running and (further improved) forthcoming
experiments, I attempt a general analysis of
the fate of Lorentz symmetry in quantum spacetime.
In particular, I analyze the deformation of Lorentz symmetry
that holds in certain noncommutative spacetimes
and the way in which
Lorentz symmetry is broken in other noncommutative spacetimes.
I also observe that discretization of areas (and/or lengths/volumes/times)
does not necessarily require departures from Lorentz symmetry,
just like the discretization of angular momentum in ordinary
quantum mechanics does not require departures from
space-rotation symmetry. This is due to the fact that
Lorentz symmetry has no implications for exclusive measurement of
the area of a surface, but it governs the combined measurements
of the area and the velocity of a surface. In a quantum-gravity theory
Lorentz symmetry can be consistent with area discretization, but only
when the observables ``area of the surface" and ``velocity of the
surface" enjoy certain special properties.
I argue that the status of Lorentz symmetry in the
loop-quantum-gravity approach requires careful scrutiny, since areas are
discretized within a formalism that, at least presently, does
not include an observable ``velocity of the surface".
In general it may prove to be very difficult to reconcile Lorentz symmetry
with area discretization in
theories of canonical quantization of gravity,
because a proper description of Lorentz symmetry appears to
require that the fundamental/primary
role be played by the surface's world-sheet, whose ``projection" along the
space directions of a given observer describes the observable area
(just like the observable ``$L_x$" is
the projection of the angular-momentum, a
legitimate ``space-vector observable" of nonrelativistic quantum mechanics,
along the ``$x$" axis of an observer),
whereas the canonical formalism only allows the introduction as primary
entities of observables defined at a fixed (common) time,
and the observers that can be considered must share that time variable.
I also comment on the measurability
of lengths/areas/volumes in theories that quantize the
fields, such as the metric, that describe spacetime geometry:
for example, I show that the same conceptual ingredients
that lead to the description of the area of a surface
as a quantum operator should also motivate a reanalysis
of the operative definition of area, and, even when formally allowed,
area discretization might be unobservable, in some sense ``hidden" behind
a fundamental limit on the measurability of areas.}
}

\newpage
\baselineskip 12pt plus .5pt minus .5pt
\pagenumbering{arabic}
\pagestyle{plain}

\section{Introduction and summary}
In quantum-gravity research it is not uncommon to find
hints of some departure from ordinary classical Lorentz symmetry,
but the associated new effects, if at all present, would be
strongly suppressed by the smallness of the Planck length
($L_p \equiv \sqrt{\hbar G/c^3} \sim 1.6 {\cdot} 10^{-35}m$).
For quite some time the assumption that they would be
unobservably small has led to scarce interest in the possibility
of departures from Lorentz symmetry.
This state of affairs is changing as a result of the
recent realization (see, {\it e.g.},
Refs.~\cite{grbgac,gampul,billetal,ita,gactp2})
that the sensitivities of experimental tests of Lorentz symmetry
are reaching the point at which even Planck-length-suppressed
Lorentz-violation effects could be studied.
As the debate on Lorentz symmetry
in quantum spacetime is gaining depth, it is emerging
that the different quantum-gravity approaches
may lead to a large variety of scenarios for what concerns
the fate of Lorentz symmetry.

The study of this interesting issue is slowed down by
technical problems (in certain quantum pictures of spacetime
it is even difficult to introduce the technical tools needed
for the analysis of the rules of transformation from one
inertial observer to another)
and by a sort of language problem.
With ``language problem" here I mean that there is not even
a consensus on the questions
to be investigated in order to establish what happens to
Lorentz symmetry;
moreover, while, as mentioned, many different things happen
to Lorentz symmetry in different quantum-gravity approaches,
usually authors refer to all these scenarios using the
single (and sometimes misleading) characterization as violations
of Lorentz symmetry.

While of course the issues of interest for
physics are the ones corcerning the physical predictions
associated with Lorentz symmetry, often Lorentz transformations
are introduced only as a formal property of the technical tools
that are used at the level of formalism. The formal structure
of the theory and the nature of the physical predictions
(for what concerns Lorentz symmetry) are very simply connected
in classical spacetime, but, as I shall clarify in the following,
this connection can be more subtle in certain quantum pictures
of spacetime.

Hoping to provide useful tools for the mentioned ``language" issue, I
introduce here (in Section~5) a terminology for the description of various
scenarios for the fate of Lorentz symmetry in quantum spacetime.
I also propose (in Section~2 and 3) a careful physical/operative
definition of a classical symmetry, which applies also to theories that are
not themselves classical.

On the technical side I contribute here an analysis
of the fate of Lorentz symmetry
in certain noncommutative spacetimes and in loop quantum gravity.
The part that concerns noncommutative spacetimes is mostly a
review of recent developments, which I discuss at a rather
intuitive level (while the original analyses involve relatively heavy
mathematics). The part on loop quantum gravity is original.
Since the aspect of loop quantum gravity which I investigate
is the interplay between area/volume discretization and Lorentz symmetry,
most of my observations actually apply equally well
to loop quantum gravity and to any other possible canonical-quantization
theory of gravity which predicts area/volume discretization.

I observe (Sections~3 and 4)
that discretization of length/area/volume
observables does not in general require departures from ordinary
classical Lorentz symmetry,
just like the discretization of angular momentum in ordinary
quantum mechanics (Section~2)
does not necessarily require departures from space-rotation symmetry.
This is due to the fact that
Lorentz symmetry has no implications for exclusive measurements of
the area of a surface, but it governs the combined measurements
of the area and the velocity of a surface. In a quantum-gravity theory
Lorentz symmetry can be consistent with area discretization, but only
when the observables ``area of the surface" and ``velocity of the
surface" enjoy certain special properties.
I argue (Section~7) that the status of Lorentz symmetry in the
loop-quantum-gravity approach requires careful scrutiny, since areas are
discretized within a formalism that, at least presently, does
not include an observable ``velocity of the surface".
I also observe that, in order to allow the compatibility
of Lorentz symmetry with area discretization,
this still unknown velocity operator
should turn out to have quite a few {\it ad hoc} properties.
Moreover, some of my results suggest that these
hypothetical properties of the surface-velocity operator
might then interfere with the role that the velocity of
a surface often (always?) plays in surface-area measurements.
It appears likely that in the end it will turn out to be impossible to
have a pair of surface-area and surface-velocity
operators that
on the one hand allow the compatibility of area discretization
with Lorentz symmetry and on the other hand allow the identification
of meaningful measurement procedures that can endow with operative
meaning the area-discretization prediction but are not affected
by an in-principle measurability limit (which would render
discreteness unobservable, spoiling it of its tentative
status as a physical prediction).
I analyze a couple of measurement procedures that had been previously
considered as possible ways to render operatively meaningful
the concept of loop-quantum-gravity area discretization and I find
that indeed (if the surface-velocity observable has the properties
required to render area discretization compatible with lorentz
symmetry) they are affected by an in-principle measurability
limit which renders area discreteness unobservable.

I also formulate the hypothesis that
these difficulties, if at all present (their full assessment
still requires additional, more refined, analyses),
might not be a genuine consequence of the physical and formal intuition
that motivated the loop-quantum-gravity approach, but rather
an artifact of the attempt to cast that intuition in the
framework of canonical quantization, which, in light of the
lack of ``democracy" between time and space,
might be ill suited for (an extension to the quantum realm of)
general relativity.
In general it may prove to be very difficult to reconcile Lorentz symmetry
with area discretization in
theories of canonical quantization of gravity,
because a proper
description of Lorentz symmetry appears to
require that the fundamental/primary
role be played by the surface's world-sheet, whose ``projection" along the
space directions of a given observer describes the observable area,
whereas the canonical formalism only allows the introduction as primary
entities of observables defined at a fixed (common) time,
and the observers that can be considered must share that time variable.

In my analysis (Section~6) of
some examples
of flat noncommutative spacetimes I focus on the aspects
that are significant for establishing what happens to
Lorentz symmetry.
In particular, I discuss the {\underline{deformation}} of
Lorentz symmetry that holds in certain
Lie-algebra noncommutative spacetimes,
and the {\underline{violation}} of
Lorentz symmetry that holds in canonical
noncommutative spacetimes.

\section{Symmetries in classical and nonclassical physics}
\subsection{Physical characterization of a symmetry}
Symmetries in physics are of course a characteristic
of our observations. Certain limitations on the variety of
phenomena we observe and certain types of relations between
our observations
are what we call symmetry transformations.
At the level of the mathematical
formalisms we use to describe our observations
one represents physical symmetries through certain properties
of the mathematical tools introduced in our formalisms.
But, of course, the role of formalism is secondary.
Symmetries are a characterization of the phenomena that
we observe and a theory will enjoy those symmetries
if the processes it predicts are governed by those symmetries\footnote{Here
I am attempting to be careful in describing a ``symmetric theory"
as a theory that predicts certain types of physical
processes, rather than as a
theory whose mathematical tools enjoy certain types of properties.
Of course, in a given class of theories one can easily identify
the properties that the mathematical tools must enjoy
in order to predict the types of processes required by a given symmetry.
However, it is dangerous to then identify the
concept of symmetry with a given set of mathematical properties:
as one moves from one class of theories to another ({\it e.g.}
from theories in commutative spacetime to theories in noncommutative
spacetime)
it is not unplausible
that the same mathematical properties of some of the technical
tools that compose the theory would lead, for example, to
different relations between the processes predicted by the theories.
As I intend to emphasize in this paper,
this is a delicate issue especially for theories
involving a nonclassical picture of spacetime.}.


The symmetries of interest in this paper are space
or spacetime symmetries, namely space-rotation
symmetry and Lorentz symmetry.
At the most fundamental level these symmetries
are characterized by the associated
symmetry transformations which
describe how the same physical process appears
to different observers.
It is at that level that one is really forced to consider
space-rotation transformations and Lorentz transformations:
the laws of physics describe the processes that can occur,
and an important aspect of these processes
is that they are ``real"/``objective",
{\it i.e.} they are observed (in principle) by all observers,
so there must necessarily be some rules (which are also to be seen
as laws of physics) that describe how the same physical process
appears to different observers.

At least for what concerns the present study the concept of ``symmetry
transformations" allows a rather intuitive description:
it pertains the objectivity of certain entities, in spite of the fact
that different observers may obtain different sets
of measurement results in their characterization of these
entities. For example, in theories that are space-rotation invariant
one can introduce the objective/physical concept of angular-momentum
vector. Different observers (observers with different orientation of
their axes) will describe this objective vector in terms of different
triples of measured quantities (the components of the angular
momentum). If the triples of two different observers concern
the same objective angular-momentum vector they must be connected
by a space-rotation transformation.

Symmetry transformations
govern the way in which the same process appears to
different observers, but in turn the presence of
some symmetries imposes constraints on the laws
of physics that apply to each of these different observers.
For example,
in a physical world with space-rotation symmetry
the fact that two different observers must make sense,
in the way prescribed by space-rotation symmetry,
of their common observations also imposes, by logical
consistency, some corresponding constraints on the laws of physics
that apply to each observer ({\it i.e.} on the processes
that each observer can observe).
These constraints are ``space-rotation symmetry
of the laws of physics written by each observer".
In sloppy but intuitive language one could say
that space-rotation symmetry
has two roles: (1) it governs how
the same physical process appears to different observers
and (2) it imposes constraints on the processes
observed by each observer.

A physical characterization of these different roles of
space-rotation symmetry will be provided in the next Subsection.
Here let me be concerned with the fact that
it can at times create confusion to use the same name for
these different roles of symmetries.
The two roles of the concept of symmetry are directly
interconnected (one follows logically form the other
and vice versa) but it is nevertheless useful to keep
clearly separate the different logical role of the two concepts.

The first concept of symmetry pertains the description of
how the same physical process appears to different observers.
I shall refer to this as the case of a ``{\underline{passive}} symmetry
transformation",
to emphasize that it involves a single physical process and the
symmetry transformations describe how that physical process appears
to different observers.

The second concept of symmetry pertains the structure of
the laws of physics that govern the physics as seen
by each of the observers. It restricts the class of processes
that any given observer can witness and it also governs
certain connections between the different processes that
a given observer can witness.
I shall refer to this as the action
of ``{\underline{active}} symmetry transformations",
to emphasize that it pertains different processes observed
by a single observer.

As mentioned, these two concepts are intimately related.
However, I will give in this paper priority to
passive symmetry transformations (reflecting an intuition
that they might be in some sense more fundamental at the conceptual
level:  it is absolutely necessary to have
some rules that govern how the same physical
process appears to different observers).

\subsection{Space-rotation symmetry in classical mechanics}
Let me characterize space-rotation symmetry through an explicit example.
When one observer measures one or more components of the angular momentum
of a classical system, using space-rotation symmetry
some facts can immediately be deduced about how that same angular
momentum appears to a second observer, whose reference axes are rotated
with respect to the ones of the first observer.
This is basically not much more than a statement
that physical processes are real/objective (in the sense intuitively
introduced above), that space does not have preferred directions and that
there is no preferred observer.

Let us call $(x,y,z)$ the axes of the first observer $O$
and $(x',y',z')$ the axes of the second observer $O'$.
Focusing, for simplicity,
on the example of angular momentum,
I found useful to note here
some characteristic properties of space-rotation symmetry:

$\bullet$ A physical entity whose objectivity is
codified in space-rotation symmetry transformations is the angular-momentum
vector. This physical entity has the primary role both in measurement
and in theory. However, each observer characterizes this vector
by three (real, dimensionful) measured numbers. Each of these numbers
is to be seen as the projection of the objective vector along
one of the axes of the observer, and, of course, in turn these axes
must be physically identified by the observer. For example, an observer
may choose as ``$x$ axis" the direction of a certain magnetic field,
another vector, and in that case a crucial role is played by
the fact that both in measurement and in theory one can meaningfully
consider the projection $\vec{L} {\cdot} \vec{B}$.
The observable simply denoted by ``$L_x$" in the formalism
inevitably corresponds physically to an observable obtained from
two objective vectors, the angular-momentum vector $\vec{L}$ and
a second vector such as $\vec{B}$. When the value of $\vec{B}$
is known one can set up a measurement procedure
for $L_x \equiv \vec{L} {\cdot} \vec{B}$.

$\bullet$ When $O$ measures the $x$ component of the angular momentum
it is still not possible to predict the components
of that angular momentum along the $(x',y',z')$ axes of $O'$,
but of course the $x$ direction is also meaningful for $O'$
and that information is acquired also by $O'$. [For example,
if the $z$ and $z'$ axes coincide and an angle $\alpha$ characterizes
the rotation of $(x',y')$ with respect
$(x,y)$, then $O'$ describes the measurement done by $O$ as a
measurement of the component of angular momentum
along the direction $cos(\alpha) \vec{x'} + sin(\alpha) \vec{y'}$.]

$\bullet$ When $O$ measures all three components, along
the $(x,y,z)$ axes, of the angular momentum
{\underline{everything}} can be said about all of the components
of that angular momentum along the $(x',y',z')$ axes of $O'$.
The triads $(L_x,L_y,L_z)$ and $(L_{x'},L_{y'},L_{z'})$
are of course {\underline{different}} but they are related by a simple
rule of transformation (a space-rotation transformation).

$\bullet$ When $O$ measures the modulus of the angular momentum
{\underline{everything}} can be said about how that modulus
appears to a second observer:
the value of the modulus is {\underline{the same}}
for both observers.

These remarks have to do with the passive space-rotation symmetry.
The active symmetry (again in the angular-momentum sector)
is encoded in other related properties. For example,
in a world with space-rotation symmetry a collection of systems
prepared in a way that does not break that symmetry will have
to enjoy, as an esemble, the same properties along any given
direction ({\it e.g.} the ensemble of measurements of $L_x$
will have to give the same result as the ensemble of measurements
of $L_y$).
Another example of manifestation of space-rotation symmetry
within the class of processes observed by a single observer
is the fact that the total angular momentum of an isolated
system does not change in time (space-rotation symmetry
imposes a constraint on the physical processes observed
by a single observer by disallowing processes in which the
total angular momentum of an isolated system is not a constant
of time evolution).

With time physicists have learned that
these physical properties of active space-rotation-symmetry
transformations and passive space-rotation-symmetry
transformations are predicted by mathematical theories
that involve the angular momentum vector in a certain
technical way ({\it e.g.} relying on Hamiltonians that
enjoy the technical/mathematical property
which is known as ``invariance under space-rotation transformations").

\subsection{Space-rotation symmetry in nonrelativistic
quantum mechanics}
Space-rotation symmetry is a classical continuous
symmetry.
Being a classical symmetry it may appear not obvious that
it can be preserved upon quantization. However, ordinary
non-relativistic quantum mechanics ``lives" in the same spacetime
as classical non-relativistic quantum mechanics.
One quantizes the entities that ``live" in spacetime,
but {\underline{spacetime is still classical}}.
It is therefore not unplausible (and not even surprising)
that one might introduce
new, non-classical, rules of mechanics without modifying the
classical space-rotation symmetry.
One might, at first sight, be skeptical that
some rules of mechanics that discretize angular momentum
could enjoy a continuous symmetry, but more careful reasoning
will quickly lead to the conclusion that there is no {\it a priori}
contradiction between discretization and a continuous symmetry.
In fact, as I emphasize below, the type of
discretization of angular momentum
which emerges in ordinary non-relativistic quantum mechanics
is fully consistent with classical space-rotation symmetry.

This concept of a non-classical ({\it e.g.} quantum) theory that
enjoys classical symmetries will be more carefully introduced
in Subsection~2.5.
The point will be that I propose to accept that a quantum
theory enjoys the classical symmetries when all the
measurements that the quantum theory still allows
are still subject to the rules imposed by the classical symmetry.
Certain measurements that are allowed in classical mechanics
are no longer allowed in quantum mechanics, and on those measurements
it will of course be impossible to verify the validity of a symmetry
in a quantum-mechanical theory; however, this does not amount to
a violation of the classical symmetry, but merely to a reduction
in the number of testable predictions of the symmetry.
For the scopes of my analysis, pertaining to classical symmetries,
it is convenient to reserve the term ``measurement"
to the situation in which the information extracted from the
system is essentially classical, so in quantum mechanics
I will be focusing on eigenstates of the observable under
consideration.
Besides the action of the classical symmetry on eigenstates,
in some cases
(when not dealing with eigenstates or
when concerned with more than one observable,
not mutually commuting)
I will also consider the action of the classical symmetry
on the expectation values of relevant observables.
Again I will insist that on these
expectation values the classical symmetry acts just as it acts
on expectation values in classical contexts in which (in spite
of the lack of an in-principle obstruction) one ends up not
acquiring full information on the observables of interest.

This definition of classical symmetry
applies in particular to certain familiar
systems of ordinary quantum mechanics
which enjoy classical space-rotation symmetry (these systems are already
described in the literature as space-rotation symmetric systems).
My definition of classical symmetry will allow me, in the
later sections,
to differentiate between quantum pictures of spacetime that do
enjoy certain classical spacetime symmetries
and pictures that do not
(not even in the sense that certain systems of ordinary quantum mechanics
enjoy space-rotation symmetry).

It is useful to note here certain properties that characterize
the presence of classical space-rotation symmetry in ordinary nonrelativistic
quantum mechanics:

$\bullet$ As in classical mechanics,
space-rotation symmetry transformations endow the angular-momentum
vector $\vec{L}$ with physical reality/objectivity.
The formalism refers most primitively to $\vec{L}$ and it makes connection
with the components of $\vec{L}$ which observers can measure
through projections completely analogous to the ones relevant
in classical mechanics, such as $L_x \equiv \vec{L} {\cdot} \vec{B}$.

$\bullet$ Just as in classical mechanics,
when $O$ measures the square-modulus  $L^2$
of the angular momentum, {\underline{everything}}
can be said about how that square-modulus
appears to a second observer:
the value of the modulus is {\underline{the same}}
for both observers. It happens to be the case that the values of $L^2$
are constrained by quantum mechanics on a discrete spectrum,
but this of course does not represent an obstruction
for the action of the continuous symmetry
on invariants, such as $L^2$.

$\bullet$ When $O$ measures the $x$ component, $L_x$, of the angular momentum
it is still not possible to predict the value of
any of the components
of that angular momentum along the $(x',y',z')$ axes of $O'$.
This is true at the quantum level just as much as it is
true at the classical
level. This is another example of situation in which
the fact that quantum mechanics constrains
the values of an observable, $L_x$, on a discrete spectrum is irrelevant
for our symmetry considerations, since the relevant symmetry
does not prescribe how that same observable is seen by another
observer.
(Note that if another mechanical theory, clearly different
from quantum mechanics, allowed simultaneous eigenstates
of ${\hat{L}}_x$, ${\hat{L}}_y$, ${\hat{L}}_z$
and predicted discrete spectra for all of them,
then the classical continuous space-rotation symmetry would
inevitably fail to apply.)

$\bullet$ Let me make one more remark on the case
in which $O$ measures the $x$ component, $L_x$, of the angular momentum.
Although nothing can be said about any of the components
of that angular momentum along the $(x',y',z')$ axes of $O'$,
of course the $x$ direction is also meaningful for $O'$
and that information is acquired also by $O'$. For example,
the fact that a system is in an eigenstate of ${\hat{L}}_x$
(the component of ${\vec{L}}$ along the $x$ axis of a certain observer)
is an objective fact that affects the observations on that system
by all observers,
although for some observers it will require a complicated description
in terms of their natural axes. For example,
if the $z$ and $z'$ axes coincide and an angle $\alpha$ characterizes
the rotation of $(x',y')$ with respect
$(x,y)$, then the observer $O'$ would describe an eigenstate
of ${\hat{L}}_x$ as an eigenstate of the component of angular momentum
along the direction $cos(\alpha) \vec{x'} + sin(\alpha) \vec{y'}$.
The fact that the statement "the system is in an eigenstate
of ${\hat{L}}_x$"
must be true for all observers (or true for none) is obvious logically.
Think for example of a beam of particles prepared in certain ways
and then sent through a Stern-Gerlach device: if one observer
sees that the way in which the beam was prepared selects eigenstates
along a certain specific direction identified by the Stern-gerlach
device, then all other observers will have to agree on that statement
(they will also see the corresponding special behaviour of that
beam going through the Stern-Gerlach device, although they might
describe it in a slightly more complicated way when making
reference to their own preferred axes of reference).

$\bullet$ In classical physics space-rotation symmetry
also governs the relation between the triple
measurement $(L_x,L_y,L_z)$ made by $O$ and the corresponding
measurement of $(L_{x'},L_{y'},L_{z'})$ made by  $O'$.
[More precisely it imposes that when $O$ attributes to
a system angular momentum with components $(L_x,L_y,L_z)$
then $O'$ assigns to that same system angular-momentum components
$(L_{x'},L_{y'},L_{z'})$, where $(L_{x'},L_{y'},L_{z'})$
is the appropriate ($O \rightarrow O'$) rotation of $(L_x,L_y,L_z)$.]
This statement is neither true nor false in quantum mechanics.
In fact, quantum mechanics excludes the possibility of
simultaneous classical/sharp measurement of all components
of angular momentum\footnote{Of course, only the properties of
generic eigenstates are of interest here.
The fact that one could have an eigenstate with $L_x=L_y=L_z=0$,
in the special case $L^2 = 0$, has no implications for my argument.
Also note that the condition $L_x=L_y=L_z=0$ does not involve
the discretization scale $\hbar$ and is space-rotation
invariant both at the classical and the quantum level
($L_x=L_y=L_z=0$ $\rightarrow$ $L_{x'}=L_{y'}=L_{z'}=0$).}.
This prediction of the classical symmetry
is not verifiable in ordinary quantum mechanics, but it would
be improper to say that it fails. My definition of a classical
symmetry in nonclassical theories will allow for these situations:
the theory can still enjoy a classical symmetry even though
some of the predictions of the symmetry cannot be tested because
of in-principle obstructions present in the nonclassical theory.
However, for the predictions that can be tested there cannot
be ``anomalies": the classical symmetry
will hold in the nonclassical theory only if all of its predictions that
are still testable turn out to be still fully successful (and
successful in the same sense as they are in the corresponding classical
limit). Moreover, a necessary condition for space-rotation
invariance is that on expectation values the symmetry must behave
as expected: in a generic state of the system (or ensemble of systems)
the expectation
values $(<{\hat{L}}_{x'}>,<{\hat{L}}_{y'}>,<{\hat{L}}_{z'}>)$
should be related to the expectation
values $(<{\hat{L}}_{x}>,<{\hat{L}}_{y}>,<{\hat{L}}_{z}>)$
by the relevant space rotation. This is indeed
what happens in those ordinary
quantum-mechanical frameworks which are
described in the literature as enjoying space-rotation symmetry.

$\bullet$ One final remark that I find useful to make on
the space-rotation classical symmetry of certain ordinary
nonrelativistic quantum-mechanical systems concerns
the role that the active space-rotations have
in those theories. All previous remarks focused on ``passive
symmetry transformations", {\it i.e.} how the same {\underline{single}}
process/property appears to {\underline{different}} observers.
Now let me stress that
a {\underline{single}} observer can of course test the presence
of (active) space-rotation symmetry.
In particular, just as in happens at the classical level,
the total angular momentum of an isolated system is
a constant of motion within ordinary quantum mechanics
(the commutator between the Hamiltonian and the total angular
momentum is zero). As another example of an ``active" role
for space-rotation symmetry,
let us consider the Stern-Gerlach
device and imagine that an observer performs the first
ever Stern-Gerlach experiment on a beam prepared in such a way
that it has no preferred axis. That observer finds evidence
of a discrete spectrum of a specific form (in that case a very simple form).
A necessary condition for (active) space-rotation symmetry
is that upon  repeating the experiment after rotating arbitrarily
the Stern-Gerlach device the same type and form of discretization
is found again. The fact that a single observer
cannot identify a preferred direction is a necessary
condition for (active) space-rotation symmetry.
This condition is clearly satisfied in
ordinary quantum-mechanical systems.
The careful reader will easily deduce the simple
relation between these properties under active symmetry
transformations and their counterparts for ``passive"
symmetry transformations.

\subsection{More on the $(L_{x},L_{y},L_{z})$ measurement}
The characterization of space-rotation symmetry
within ordinary quantum mechanics provided in the previous Subsection
is sufficient for the purposes of the present analysis.
However, my characterization of a classical symmetry (inside
and outside classical physics) intends to be useful also for
future studies, particularly future studies considering the
fate of classical Lorentz symmetry in quantum spacetime.
In this respect it may prove useful to pause here for
a more careful analysis of the triple measurement $(L_x,L_y,L_z)$
within classical space-rotation symmetry.

In this Subsection I focus, for simplicity,
on two observers $O$ and $O'$ with common orientation
of the $z$ axis and with a relative angle $\alpha$ rotation
of the axes on the $x,y$ ($x',y'$) plane.

As mentioned, within classical physics space-rotation symmetry
transformations govern the map between
a measurement $(L_x,L_y,L_z)$ made by $O$ and the corresponding
measurement of $(L_{x'},L_{y'},L_{z'})$ made by  $O'$.
A classical beam of classical particles which is prepared
in such a way that they all have the same values of $L_x$, $L_y$, $L_z$
when studied with respect to axes $x',y',z'$ (the natural axes
of observer $O'$) will be found to be a beam in which all particles
have the same values of $L_{x'}$, $L_{y'}$, $L_{z'}$,
with $(L_{x'},L_{y'},L_{z'})$ being the appropriate rotation
of the original triple $(L_x,L_y,L_z)$,
which for the chosen pair of observers
satisfies $L_{x'} = cos(\alpha) L_x + sin(\alpha) L_y$,
$L_{y'} = - sin(\alpha) L_x + cos(\alpha) L_y$,
$L_{z'} = L_z$.
Since these rotation maps are continuous, by changing the
relative orientation of the axes of $O$ and $O'$ one finds values
of $L_{x'}$ (and of course the same holds for $L_{y'}$)
that take arbitrary value within a relevant range.

As emphasized above,
ordinary quantum mechanics is an example in which
this aspect of space-rotation symmetry cannot be tested.
The simultaneous measurement of $L_x$, $L_y$, $L_z$
is not allowed by the laws of quantum mechanics.
It is not through
this triple measurement $(L_x,L_y,L_z)$ that one
can find space-rotation symmetry to fail (or succeed).
However, the logical structure of the role of space-rotation
symmetry (especially since it is a continuous symmetry)
in quantum mechanics is manifest in the specific
way in which quantum mechanics imposes limitations on the
simultaneous measurability of $L_x$, $L_y$, $L_z$.
Consider a beam of particles which have been prepared in an eigenstate
of ${\hat{L}}^2$ and ${\hat{L}}_x$ and are found
to all have $L^2 =3 \hbar/4$ and $L_x =\hbar/2$, where $x$ identifies
the $x$ axis of observer $O$.
Quantum mechanics also predicts that when $L_x$ is fully known,
the values of $L_y$ and $L_z$
must be affected by a large uncertainty.
The observer $O'$ could measure $L_{x'}$, finding that the same
beam does not correspond to an eigenstate of ${\hat{L}}_{x'}$,
but rather $L_{x'}$ takes values $\hbar/2$ and $-\hbar/2$ with a certain
probability within the beam. This probability distribution
will be characteristic of the fact that the beam is described
by an eigenstate with $L_x =\hbar/2$. For example, for small angles $\alpha$
positive values ($\hbar/2$) of $L_{x'}$ will dominate on the
negative values ($-\hbar/2$),
and overall $<{\hat{L}}_{x'}> = cos(\alpha) \hbar/2$.

In the narrow context here considered (and only in a handful
of similar contexts) one could roughly
describe in classical-physics language the prediction
of ordinary quantum mechanics.
Ordinary quantum mechanics roughly states that a beam of particles
can be characterized by a common value of, say, $L^2$ and $L_x$ but then
inevitably $L_y$ will vary within the beam in totally random
manner. Eigenstates of ${\hat{L}}^2$,${\hat{L}}_x$
with $L^2 =3 \hbar/4$ and $L_x =\hbar/2$
are states in which $L_y$ is undetermined but $<L_y>=0$ and
$\sigma_{L_y}^2 \equiv <L_y^2> - <L_y>^2= \hbar^2/4$.
This point clearly plays a key role in the consistency
between the discretization of angular momentum predicted
by quantum mechanics and the space-rotation symmetry
of quantum mechanics. For example even for small $\alpha$ in a rich
beam (a beam with an infinite number of particles) there will be some
small percentage of particles whose $L_{x'}$ is found to be
negative, $L_{x'} =-\hbar/2$.
Since $L_{x'} = cos(\alpha) L_x + sin(\alpha) L_y$, $<L_y>=0$,
and\footnote{Here I am taking abundant liberty in adopting
a classical language, but the probabilistic considerations, and the
role of the $\sigma_{L_y}$ uncertainty in those consideration,
are appropriate.}
$L_x =\hbar/2$ the small percentage of particles found to
have $L_{x'} =-\hbar/2$ are manifestation of the
quantum-mechanical probability distribution which
is strongly characterized by $\sigma_{L_y}=\hbar/2$ (although it
is far from being a gaussian: it assigns nonvanishing probabilities
only at $L_{x'} =\hbar/2$ and $L_{x'} =-\hbar/2$).
In this entire probabilistic description, which is at the core of
the workings of space-rotation symmetry in ordinary quantum mechanics,
an important role is clearly played by the size of $\sigma_{L_y}$.
If $\sigma_{L_y}$ had taken value, say, $\sigma_{L_y}=\hbar/1000$
it would have not been possible for $<{\hat{L}}_{x'}>$
to take the value $<{\hat{L}}_{x'}> = cos(\alpha) \hbar/2$.

This allows us to deduce that in order for discretization to
be compatible with continuous classical symmetries
it is not only necessary that combinations of observables
governed by the symmetry, like $(L_x,L_y,L_z)$,
should not be measurable simultaneously: it is also necessary
that the eigenstates of one of the relevant observables,
say ${\hat{L}}_x$, be characterized by an appropriately
large uncertainty in the other relevant observables
(${\hat{L}}_y$ and ${\hat{L}}_z$).

\subsection{Classical symmetries in any (classical or non-classical)
theory}
The observations reported in the preceding Subsections are the
basis for my definition of the presence of a classical symmetry
in a non-classical, {\it e.g.} quantum, theory.
The role that classical
space-rotation symmetry plays in certain
contexts of ordinary nonrelativistic quantum mechanics
will be my prototype for the role that a classical symmetry
should play in a non-classical theory in order for us to
state that the symmetry holds.
The fact that my definition of classical symmetry
applies (by construction) to the case of space-rotation symmetry
in ordinary quantum mechanics assures me of the
fact that the definition is not purposeless. In contexts in
which a classical symmetry characterizes observations
governed by a nonclassical theory ``less strongly" than
in the case of space-rotations in ordinary quantum mechanics
it is appropriate to state that the classical symmetry
is (perhaps partly or softly) violated: we should reserve
the name ``classical symmetry" to contexts in which the symmetry
characterizes observations as strongly as in the remarks
made in the Subsection~2.3.

By stating that the role
that classical space-rotation symmetry plays in certain
contexts of ordinary nonrelativistic quantum mechanics
is my prototype for the role that a classical symmetry
should play in a non-classical theory
I have provided a definition which should be
clear to the careful reader.
It is nevertheless useful to stress here some of the points
that emerged in the preceding subsections.

The basic point is that the operation of measuring one, two
or more observables will always end up giving some main
estimate of the observables and some uncertainties.
For example, in the case of two observables, $R$ and $S$,
the measurement result would be
of the type $R = R_0 {\pm} \delta R$, $S = S_0 {\pm} \delta S$.
The action of the classical symmetry should not be affected
by the nature of the uncertainties $\delta R$, $\delta S$:
the classical symmetry acts in the same way independently of
whether the uncertainties are ``fundamental"
(due to a quantum-mechanical uncertainty principle)
or due to technological/practical limitations.
The most significant features of the classical symmetry
emerge by considering the case $\delta R = \delta S= 0$,
which is at least available (in principle)
in the classical-theory limit.
If the ($R$,$S$) measurement is meaningful for the symmetry,
when a given ($R$,$S$) measurement procedure on a given system
gives result $\delta R = \delta S= 0$, $R = R_0$, $S = S_0$
(a ``sharp" measurement)
for observer $O$, that same\footnote{The reader will be
in a position to appreciate fully
the strong sense in which I intend the statement ``that same
measurement procedure on that same system" after reading
Subsections~3.3, 3.4 and 3.5.} measurement procedure on that same
system should
give result $\delta R' = \delta S' = 0$, $R = R_0'$, $S = S_0'$
for observer $O'$, where ($R_0'$,$S_0'$)
is related to ($R_0$,$S_0$) by the relevant $O \rightarrow O'$
symmetry transformation.
If according to $O$ the measurement procedure is affected
by non-zero uncertainties $R = R_0 {\pm} \delta R$, $S = S_0 {\pm} \delta S$,
then according to $O'$ the measurement procedure still gives result
with $R = R_0'$, $S = S_0'$ but of course also $O'$ finds
non-zero uncertainties $\delta R'$, $\delta S'$.
$\delta R'$,$\delta S'$ is related to $\delta R$,$\delta S$
in a way affected by the structure of the symmetry transformations
but the relation is not independent of the theory one
is considering (in fact, as clarified in Subsection~2.4,
the relation also depends on the structure
of the probability distributions attributed to uncertainties
in the theory), whereas of course the relation between ($R_0'$,$S_0'$)
and ($R_0$,$S_0$) is fully specified by the symmetry, independently
of whether the theory is classical or non-classical.

It will be easy for the careful reader to verify that the properties
described in the previous long paragraph are satisfied by classical
space-rotation symmetry both in classical and in quantum mechanics.
The properties I stated apply if, as mentioned,
the ($R$,$S$) measurement is ``meaningful for the symmetry".
Of course, I describe as meaningful for the symmetry a measurement
for which the symmetry makes definite predictions.
The measurement of $L^2$ and the measurement of $(L_x,L_y,L_z)$
are examples of measurements that are meaningful for space-rotation
symmetry, while the measurements in which one only measures $L_x$
are not meaningful
for space-rotation symmetry. As clarified in the preceding Subsections,
it is not an accident that the $L^2$ measurement can be sharp (both
in classical and) in quantum mechanics, since $L^2$ is an invariant
of space rotations and its discretization will therefore not interfere
with continuous space-rotation symmetry transformations.
As also clarified in the preceding Subsections,
it is not an accident that the $(L_x,L_y,L_z)$ measurement cannot be
sharp in quantum mechanics, since $(L_x,L_y,L_z)$ is not an invariant
of space rotations and its discretization would have interfered
with continuous space-rotation symmetry transformations.
The fact that the $L_x$ measurement can be sharp in quantum mechanics
of course bears no relevance for the fate of classical space-rotation
symmetry, since the measurement of $L_x$ is not meaningful
for space-rotation symmetry: the knowledge of  $L_x$ does not allow
to establish anything about $L_{x'}$, $L_{y'}$ and $L_{z'}$,
independently of whether or not space-rotation symmetry is present.

Although the careful reader may find it redundant, for the benefit
of leasurly readers let me stress a point about eigenstates (which
is however already implicit in the remarks provided above and is therefore
indeed redundant).
If the ($R$,$S$) measurement is meaningful for the symmetry
and the theory allows the sharp measurement of ($R$,$S$)
({\it i.e.} the theory allows $\delta R = \delta S= 0$),
then the symmetry predicts without room for arbitrariness that
a system measured to have
$\delta R = \delta S = 0$, $R = R_0$, $S = S_0$
for observer $O$ must have
$\delta R' = \delta S' = 0$, $R = R_0'$, $S = S_0'$
for observer $O'$.
Otherwise the non-classical theory would be allowed to violate a
prediction of the classical symmetry:
two observers would be analyzing the same measurement
procedure and find results for ($R$,$S$)
that are not directly connected by the symmetry.
This should not be allowed if the classical symmetry does hold
in the non-classical theory, and in fact it does not
happen in ordinary quantum mechanics, where classical space-rotation
symmetry does hold.
In the language of quantum mechanics
this can be described with the statement that ``eigenstates
of a combination of observables ($R$,$S$)
which is meaningful for the symmetry must be mapped by the symmetry
into states which are ($R'$,$S'$) eigenstates",
as indeed it happens to eigenstates of $L^2$ in ordinary
quantum mechanics.

On all measurements that can be performed in
a ``classical sense" ({\it e.g.} by preparing/observing
a suitable eigenstate, or a suitable ensemble of eigenstates)
the symmetry acts just as in the classical limit: the relation
between the values assigned to observables of a given system
by two different observers is governed by the classical symmetry.
The nonclassical theory can limit
the types of ``classical measurements" that can be performed
({\it e.g.} in quantum mechanics the simultaneous classical/sharp
measurement of $L_x$, $L_y$, and $L_z$ is excluded).
This will not be described as a failure of the classical
symmetry; however, in these cases
the presence of the classical
symmetry should be reflected at the level of expectation values.
This is here stated in the same sense that in ordinary
quantum mechanics, as emphasized in
Subsection~2.3, classical space-rotation symmetry does connect
the expectation values $(<L_{x'}>,<L_{y'}>,<L_{z'}>)$
and the expectation values $(<L_{x}>,<L_{y}>,<L_{z}>)$
in a generic state of the system.

The analysis reported in the preceding subsections
allows us to describe in general terms what are the conditions
for compatibility between the presence of a
classical {\underline{continuous}}
symmetry (as here defined) and
the emergence of ``{\underline{discrete}} spectra"
(the case in which the nonclassical theory predicts that the
outcome of certain measurement procedures, the ones providing
the operative definition of one of the mathematical ``observables"
in the formalism, can only take certain discretized values).
{\bf Discretization is consistent with the classical continuous
symmetry} when it concerns observables
which are invariants of the classical-symmetry transformations,
and, of course, also when it concerns observables on which
the continuous classical symmetry makes no prediction.
A conflict emerges only when the discretized
observable is directly governed by the symmetry and the symmetry
predicts a continuous change of that observable in going from
one observer to another.
As discussed in Subsection~2.3, $L_x$ has a discrete spectrum
in quantum mechanics, but the knowledge of $L_x$ is not governed
by space-rotation symmetry ({\it e.g.} the knowledge of $L_x$
does not allow to predict the value of $L_{x'}$ in a
classical theory with space-rotation symmetry
and still does not allow to predict $L_{x'}$
in quantum mechanics). Also $L^2$ has a discrete spectrum
in quantum mechanics and the
space-rotation symmetry does govern
the knowledge of $L^2$; however, space-rotation symmetry
prescribes $L^2 = L'^2$ which is consistent with discretization
(on $L^2$ the continuous symmetry transformations
we call space rotations act trivially, $L^2$  is an invariant).

A key point for some of the considerations that are reported in the
following is the fact that spacetime symmetries basically introduce
some objective entities,
which however lead to measurement results which are not the same
for all observers: in a space-rotation-invariant world all observers,
independently of the orientation of their respective $x,y,z$ axes,
agree on the angular-momentum vector of a given system,
but the triple of measured numbers that each observer attributes to
that angular momentum depends on the observer.
The theory and the measurement procedures must make most fundamentally
reference to the angular momentum vector, which is the objective
entity, and its components will be identified through some other
physical vectors parallel to the axes of the observer
(for example, a magnetic field).
These remarks apply equally well to classical mechanics and quantum
mechanics; the only difference is that quantum mechanics imposes
some limitations on the accuracy by which one can ``measure the
vector" (measure its three components for a given observer)
and predicts a discretization of certain measurement results.

I close here my characterization of the presence of
a classical symmetry in a non-classical theory.
My characterization focuses
on ``passive" symmetry transformations, but
the implications for active symmetry transformations
can be easily deduced.

\section{Lorentz symmetry}
\subsection{Passive Lorentz-symmetry transformations}
The discussion of space rotations, on which most of the
previous Section focused, is extremely simple (so intuitive
that some statements here reported for completeness
should have appeared obvious to most readers)
and therefore ideally suited for the introduction of
the concept of classical symmetry
in a nonclassical theory that I am advocating.
The main objective of this paper
is however an analysis of the fate of
classical Lorentz symmetry in quantum spacetime.

Such an analysis of Lorentz symmetry, could have been
seen as merely academic
until only a few years ago, but it should be now perceived
as a high-priority objective, in light of
the remarkably improved sensitivity of ongoing and forthcoming
experiments, which could be
sufficient~\cite{grbgac,gampul,billetal,ita,gactp2}
to detect even tiny,
Planck-length suppressed, deviations from ordinary Lorentz symmetry.
We even already have some tantalizing experimental
hints~\cite{gactp2},
especially in the context of certain puzzling
observations~\cite{crdata}
of ultra-high-energy cosmic rays,
which could be interpreted as manifestations of a
Planck-length induced deviation from ordinary Lorentz symmetry.

At the conceptual level, while the analysis of space-rotations
in ordinary quantum mechanics is completely elementary,
the outcome of analyses of Lorentz symmetry in quantum spacetime
is not at all {\it a priori} obvious.
The point is that both space-rotations symmetry and Lorentz symmetry
are most fundamentally properties of classical space/spacetime.
Ordinary quantum mechanics, just like classical
mechanics, lives in the arena provided by classical spacetime,
and therefore,
as long as the new rules of mechanics do not explicitly break the
spacetime symmetry (and the rules of ordinary quntum mechanics
do not), it is not surprising that the classical symmetries
of classical flat spacetime survive that type of quantization.
But quantum-gravity research is encouraging many scientists
to consider one form or another of quantization
of {\underline{spacetime itself}}, so spacetime itself changes
and one can expect that in general also its symmetry
properties will change, as I shall show to be the case
in some examples considered in this paper.

The focus on (global) Lorentz symmetry is justified by our
capability to test it. Quantum-gravity research is mostly
occupied with spacetimes which are far from being flat,
but these theories must have a zero-curvature limit
and it is that limit
which we can test most accurately in ongoing and planned experiments.
Think for example of approaches to the quantum-gravity problem
that rely on noncommutative geometry: the most interesting
formal work done on these approaches concerns non-flat spacetimes;
however, if at the fundamental level spacetime geometry
is proven to be noncommutative this should in particular
apply to the physical contexts in which we basically deal
with flat spacetime. There will be a noncommutative version
of Minkowski spacetime. The
symmetry properties of this noncommutative Minkowski spacetime
are very significant,
since they can be tested very accurately.

But let us proceed step-by-step. First we need to
list a few characteristic properties of classical Lorentz
symmetry.

$\bullet$ The entities which Lorentz symmetry endows with
objective/physical/observer-independent existence
are four-vectors (tensors,...) and world-volumes (world-lines,
world-sheets,...).
The energy-momentum four-vector is ``the same" for all observers,
although each observer will describe its components in a different
way. Lorentz-symmetry transformations will govern the relation
between the components of the energy-momentum four-vector
for different observers, basically stating indeed that those
different results of measurements actually describe the same
objective energy-momentum.
Similarly the world-sheet spanned, for example, by a physical surface
is an objective entity. A given observer (with a given space/time
foliation) can describe such a world-sheet as a collection
of equal-time surfaces, and at each time instant can attribute
to the surface a velocity $V$ and an area $A$. The collection
of ($V$,$A$) as functions of time are different for different
observers, but when they refer to the same world-sheet
(the same physical surface) Lorentz-symmetry transformations
connect the values ($V$,$A$) measured by one observer with
the values ($V'$,$A'$) measured by another observer.

$\bullet$ Composition of velocities.
Consider a particle which, according to oberver $O$,
has velocity ${\vec{V}}$.
Lorentz symmetry governs the relation between ${\vec{V}}$
and the velocity ${\vec{V}}'$
that another observer $O'$ will measure for that same particle,
if the relative $O$-$O'$ velocity is known.

$\bullet$ Time dilatation.
Consider a muon which moves at speed $V$
with respect to observer $O$,
and $O$ measures the decay time $\tau$ of the muon.
Lorentz symmetry governs the relation between $\tau$
and the decay time $\tau'$
that another observer $O'$ will measure for that same muon,
if the relative $O$-$O'$ velocity is known.

$\bullet$ Length contraction.
Consider a thin straigth bar (a collection of
particles in rigid motion)
which moves at speed $V$
with respect to observer $O$.
$O$ measures the length $L$ of the bar.
Lorentz-symmetry transformations govern the relation between $L$
and the length $L'$
that another observer $O'$ will measure for that same bar,
if the relative $O$-$O'$ velocity is known.
This remark also applies in particular to wavelengths.
It also affects in an obvious way the contraction of
areas and volumes.

$\bullet$ Kinematical thresholds.
Consider the situation in which observer $O$ has two
ideal\footnote{The way in which we establish experimentally
kinematical thresholds does not follow the schematization
adopted here for simplicity. In particular, the ideal lasers
I consider are not available to us. However, the basic point is
correctly portrayed by my simplified scheme.}
photon lasers such that the energy of the emitted photons
can be tuned with arbitrarily high accuracy over an extremely wide range
of energies.
$O$ points the two lasers one toward the other, in order to study
head-on collisions, keeps one of the lasers tuned at a fixed
small energy $\epsilon$
and increases the energy of emission of the second laser gradually
from zero up to the value $E$ (the threshold energy) for which
some production of electron-positron pairs
starts to occur.
Lorentz symmetry also governs the way in which this threshold
arrangement of the experimental setup appears
to a second observer $O'$ moving with respect to $O$
at some speed $V_0$ along the axis of the collision.
Specifically, for known $V_0$,
Lorentz symmetry governs the relation between $\epsilon$,$E$
and the corresponding energies $\epsilon'$,$E'$ that $O'$
measures as emission energies of the lasers at the given threshold
condition realized by $O$ (I am, for simplicity, assuming that only
$O$ is allowed to tune the lasers).
In addition, Lorentz symmetry predicts that,
independently of $V_0$, the product $\epsilon {\cdot} E$ will have
the same value as the product $\epsilon' {\cdot} E'$
and these products will give the square of the electron
mass: $\epsilon' {\cdot} E' = \epsilon {\cdot} E = m_e^2$.

\subsection{Active Lorentz-symmetry transformations}
The list of characteristic properties of classical Lorentz symmetry
in the preceding Subsection
mainly focused on ``passive" Lorentz-symmetry transformations
(however, the careful reader will notice that
the description of the kinematical thresholds involved both
active and passive Lorentz transformations.)
Let me mention here
a couple of examples of active roles for Lorentz symmetry.
The wavelength independence of the speed of light (and the
associated form of the photon
dispersion relation, $E^2=c^2 \vec{p}^2+ c^4 m^2$) is a
prediction associated with active Lorentz-symmetry transformations
since it is a relation
between the results of different measurements
done by a single observer (speed-of-light measurements
at different energies: all photons have the same
speed independently of energy).

Another example in which active Lorentz-symmetry transformations
play a role is the muon decay time, already mentioned above
for what concerns passive Lorentz symmetry transformations.
The fact that the same muon has different
decay times for different observers (and the transformation rules
that connect those time measurements) is a prediction
associated with passive Lorentz-symmetry transformations.
The fact that the two muons with different energy (for a given
observer $O$) will have decay times that {\underline{typically}}
(on average, see below)
differ by amounts dictated by Lorentz-symmetry
trasformations is a prediction
associated with active Lorentz-symmetry transformations.

It is particularly clear in the case of the muon decay time
that the two manifestations (active and passive) of Lorentz symmetry
are deeply and simply connected.
But the muon decay time also allows us to point out a
certain difference between active and passive symmetry transformations.
In fact, the muon lifetime has a ``statistical"
component, {\it e.g..} a muon at rest ``lives" on
average $2.2 {\cdot} 10^{-6}s$, but in an ensemble of muons at rest
some live shorter than $2.2 {\cdot} 10^{-6}s$, some live longer.
In the situation I described, an observer measuring the decay times of
two muons with different energies (and therefore different speeds),
Lorentz symmetry cannot predict exactly what is the relation between
the two decay times. However, if a single observer has a large number
of muons at energy $E$ and another large number of muons
at energy $E'$ she will be able to see that Lorentz symmetry
transformations predict accurately the relation between
the average lifetimes of the two groups of muons.
When a single muon is available and it is observed by two observers
the situation is slightly different: for one observer (which could be the
rest-frame observer) the given muon will ``live" a certain time $t^*$
which might or might not coincide with the lifetime $\tau$,
according to the other observer, if indeed
Lorentz-symmetry transformations apply,
that {\underline{same}} muon will ``live" a corresponding time ${t^*}'$.
There is of course no statistical consideration that applies to
the context of a single muon observed by two observers, while
statistical considerations do play a role when comparing
the lifetimes of two muons being observed by
a single observer.

The list of examples of instances in which active or passive
Lorentz-symmetry transformations play a role is endless.
The examples I discussed probably illustrate a wide enough
ensemble of situations. But, before turning to the fate of
Lorentz symmetry in quantum spacetime,
it is perhaps useful to describe in greater detail certain
features of passive
Lorentz-symmetry transformations.
This is done in the next three Subsections.

\subsection{Time dilatation}
In order to illustrate the way in which passive
Lorentz-symmetry transformations govern the rules
of time dilatation it is sufficient to analyze
a simple clock (for special-relativity experts
this analysis is by now a textbook exercise, but I repeat it
here since it is useful for one of the points I intend to
raise about Lorentz symmetry).
Let us consider two observers, $O$ and $O'$,
each with its own spaceship, in a situation such that
the relative position and the relative velocity of the spaceships
are both pointing in the same direction (a configuration which
is effectively one-dimensional), which the observers
choose to identify with their respective $z$ axes.
Let us then mark ``$A$" and ``$B$"
two points on $O$'s spaceships (the rest frame): $A$ is
a point on the $z$ axis, while $B$ is off of the $z$ axis
and such that the segment that joins $A$ and $B$
is orthogonal to the  $z$ axis (and therefore orthogonal
to the direction of relative motion of the two observers).
The first step is for $O$ and $O'$ to measure
the distance $\overline{AB}$, to which they will end up
attributing the same value (lenghts orthogonal to the boost
direction are unaffected by boosts).
Assume then that ideal mirrors are placed at $A$ and $B$,
so that light can bounce back and forth between $A$ and $B$.
This consitutes a ``light clock".
Assuming nothing else but the constancy of the speed of light
(postulated) time dilatation follows straightforwardly.
For observer $O$ the time interval corresponding to each
tick of the light-clock is $\tau =2 \, \overline{AB}/c$.
For the second observer, $O'$, the light clock is moving. The speed of
the light used by the light clock is the same for the two observers
but the distance travelled between ticks has different values for the
two observers: that distance is $2 \, \overline{AB}$ for
the first observer, while
for the second observer it has value\footnote{Note that for observer $O'$
the light clock (which is at rest with respect to $O$) is moving
with velocity $V$, which of course coincide with the $OO'$ relative
velocity. While $O$ sees the trajectory of the light beam as going
back and forth along a straight light, $O'$ describes the
trajectory of the light beam as a ``zig-zag": for example, when bounced
back from $B$ toward $A$ the light beam, according to observer $O$,
goes in an oblique direction, and while the light beam progresses
toward $A$, the point $A$ keeps moving (it is at rest with respect
to $O$ but moves with velocity $V$ with respect to $O'$).}
$2 \, \overline{AC}/\sqrt{1 - V^2/c^2}$.
We conclude that, whereas the first (rest) observer attributes
to each tick of the light clock a time interval $\tau =2 \, \overline{AC}/c$,
the second observer attributes
to each tick of the light clock a time
interval $\tau' =2 \, \overline{AC}/\sqrt{c^2 - V^2}
=  \tau / \sqrt{1 - V^2/c^2}$,
as predicted by special-relativistic time dilatation.

A key point for the understanding of some predictions
of passive Lorentz-symmetry
transformations is that the instruments used by the first
observer
are also admissable instruments for the second observer.
Here I have discussed time dilatation using a single clock.
The two observers will of course agree on the readout of the instrument
(the objective/observer-independent number of ``ticks" done
by the light clock, which, with suitable electronics,
could correspond to a number shown by the light clock);
however, while the number of ticks of the light clock is
the same for the two observers, the time interval that the two
observers assign to each tick is different.

\subsection{Length contraction}
In order to illustrate the way in which passive
Lorentz-symmetry transformations govern the rules
of length contraction it is sufficient to analyze
a simple gedanken length-measurement procedure
(again a textbook exercise which is useful for one
of the points I want to raise about Lorentz symmetry).
Let us consider again our two observers, $O$ and $O'$,
with their spaceships. The setup is identical to the one
adopted in the previous subsection:
the relative position and the relative velocity of the spaceships
are both pointing in the same direction, which the observers
choose to identify with their respective $z$ axes,
and the previously-introduced points $A$ and $B$
coincide with the mirrors of a light clock.
In addition now let us mark a third point, ``$C$", which,
like $A$, is on the $z$ axis.
$O$ and $O'$ want to measure the distance $\overline{AC}$
(the length of a segment placed in direction parallel
to the relative $OO'$ motion).
The procedure of measurement of the distance $\overline{AC}$
is structured as a time-of-flight measurement:
an ideal mirror is placed at $C$
and the distance is measured
as the half of the time needed by
a photon wave packet, ``the probe",
sent from $A$ toward $C$ to be back at $A$
(after reflection by the mirror).
Timing is provided by the digital light-clock (involving the points $A$
and $B$) which I have already analyzed in the previous subsection.
The rest-frame observer, $O$, will measure $\overline{AC}$
as the length $L = c {\cdot} N {\cdot} \tau/2$, where $N$ is the number of
ticks done by
the digital light-clock during the $A$$\rightarrow$$C$$\rightarrow$$A$
journey of the probe. ($\tau$ is again the time interval corresponding to
each tick of the light-clock, which, as discussed in the previous
subsection, has the value $\tau =2 \, \overline{AC}/c$.)

Again, also in this more complex measurement procedure,
measuring a distance,
it is worth emphasizing (since it is
a key point for the understanding of passive Lorentz-symmetry
transformations) that the instruments used by
observer $O$ (the one on the rest-frame spaceship)
are also admissable instruments for the second observer:
the light gun is actually described in the same way
by the two observers since, according to Lorentz symmetry,
the speed of the emitted photons
is independent of the speed of the emitting gun; moreover,
as shown in the previous subsection,
an accurate light clock at rest for observer $O$ is
also an accurate moving clock for observer $O'$.
So the second observer, $O'$ can ``look at" the measurement procedure
adopted by the first observer and adopt it as its own measurement
procedure. As already emphasized in the previous subsection,
in looking at this same measurement procedure
the two observers will of course agree on the number of ``ticks" done
by the light clock during the probe's two-way journey.
That number of ticks
is an objective fact (possibly a number shown by the light clock
when triggered to stop upon the return of the probe at $A$).
Of course, while the number of ticks of the light clock is
the same for the two observers, the time interval that the two
observers assign to each tick is different, and in fact we found
in the previous subsection that according to the
second observer, $O'$, each tick of the light clock
corresponds to the time interval $\tau' =2 \,
\overline{AC}/\sqrt{c^2 - V^2} =  \tau / \sqrt{1 - V^2/c^2}$.

The other aspect of the measurement procedure that takes different
form for the two observers is the relation between the time needed
by the probe for its two-way journey and the length of the bar.
The first observer sees the bar at rest, so she uses the
relation $T = N \tau = 2L/c$.
The second observer sees the bar moving with velocity $V$
and the two parts of the two-way journey of the probe are,
for the second observer, of different length\footnote{For part
of the journey of the probe the fact that the bar is moving
shortens the probe's trip toward the next extremity of the bar,
for the other part of the journey the opposite occurs.}:
one part has length $cL'/(c-V)$
and the other part has length $cL'/(c+V)$, where I denoted
with $L'$ the distance $\overline{AC}$ according to observer O'.
So for the second observer the relation between the time needed
by the probe for its two-way journey and the length of the bar
takes the form
\begin{equation}
T' = N \tau' = {L' \over c-V} + {L' \over c+V} = {2 c L' \over c^2 - V^2}
~.
\label{relazo2}
\end{equation}
Using the relation between $\tau$ and $L$
and the relation between $\tau'$ and $\tau$ derived above
this leads to
\begin{equation}
L'= {c^2 - V^2 \over c} N {\tau' \over 2}
= \sqrt{1 - {V^2 \over c^2}} L
~.
\label{dsfitzlob}
\end{equation}
This is length contraction.
It is often said that the same distance has different value
for different observers, it is contracted in the boosted frame
with respect to its value in the rest frame.
Some authors appear to assume that this
exclusively means that
two {\underline{different}} measurement procedures, one done by
observer $O$ and another one done by observer $O'$,
would give the different results for the distance/length.
As shown in this subsection, FitzGerald-Lorentz contraction
has a stronger implication: the {\underline{same}} measurement procedure
is witnessed by the two observers, giving the same experimental
readouts (such as the readout of the light clock in my examples),
but the analysis of the measurement is different for
the two observers and leads to different conclusions
about the value of the distance/length being measured.

\subsection{Kinematic thresholds}
In the previous Subsection length contraction
was analyzed applying the concept of
passive Lorentz-symmetry transformation
in the strongest sense:
I did not just consider {\underline{the same}} length
as seen by two observers, I also considered the case
in which the two observers use {\underline{the same devices}}.
This is the sense in which passive symmetry transformations
describe how the same {\underline{measurement procedure}}
appears to two different observers.
This is, in this author's opinion, a very important aspect
of the classical symmetries under consideration in this paper:
in the analysis of situations in which two observers
share the same measurement procedure several objective
statements arise which can be of guidance for the analysis.

It is worth making another example, in addition to length contraction.
Let me look again at the kinematic-threshold procedure already
considered above and analyze it more carefully as a
measurement procedure shared by two observers.
As mentioned, I imagine that observer $O$ has two
ideal photon lasers such that the energy of emitted photons
can be tuned with arbitrarily high accuracy over an extremely wide range
of energies. The two lasers are pointed along an axis
in such a way to produce head-on collisions.
The second observer $O'$ moves with velocity $V_0$ (with respect to $O$)
along the direction of the axis of the head-on collisions.

Before starting the measurement procedure $O$ will need to
calibrate her lasers. It is not sufficient that
they can be accurately tuned, it must also be possible to establish
which energy they are emitting when tuned in a certain way.
Let us imagine that they are constructed in such a way
that the energy of the photons emitted can be tuned at any of
an infinity of energy levels, all equally spaces in energy,
so that the calibration procedure will basically amount
to establishing the $\Delta E$ gained each time that the laser
is tuned up to the next discrete level.
This calibration will be easily done by $O$ before the measurement
procedure: she will place a device that
measures the energy of photons in front of the laser,
change once from one level
of tuning to the next, and this will be sufficient for the
calibration.
I assume for simplicity that $O$ finds that both devices have the
same calibration $\Delta E$ (this will happen if the two devices
are built in exactly the same way and they are
both at rest with respect to $O$).
If $O$ and $O'$ want to share the lasers
(if they want to be able to make use simultaneously of the
same measurement procedure)
it is necessary for $O$, who has the lasers on her spaceship,
to be kind enough to send some photons from her lasers
toward some energy-measurement devices that belong to $O'$.
It is convenient for $O'$ to prepare these calibration
devices at rest
(with respect to $O'$). Moreover, $O'$ should take into account
that the two lasers point in opposite
directions on a spaceship ($O$'s spaceship) which
is moving with respect to $O'$, and therefore $O'$ should place
the calibration devices accordingly (in particular, at least one of
the calibration devices might have to be placed outside his spaceship).
$O'$ will find that the two lasers have different calibrations,
$\Delta E'_a$ and $\Delta E'_b$.
The relations between $\Delta E$ and $\Delta E'_a$,$\Delta E'_b$
are an experimental result for $O'$ but of course
we can predict them to be governed by
Lorentz transformations of energy, and depend only on $V_0$
(and the fact that one laser emits its photons with velocity
parallel to the relative $O-O'$ velocity while the other laser emits
in the opposite direction).

At this point both observers have a calibration
of the lasers and everyone is ready for the measurement procedure.
$O$ will have the task of tuning the lasers, since they
are on her spaceship (but actually $O'$ could use a remote-control
device), but everything that happens will be witnessed by both
observers.
$O$ tunes one of the lasers to a fixed level ``$n$",
which according to her calibration corresponds to
the small energy $\epsilon = n {\cdot} \Delta E$,
and increases the energy of emission of the second laser gradually
from zero up to the value $E = N {\cdot} \Delta E$ for which
some production of electron-positron pairs
starts to occur (the threshold).
The numbers $n$ and $N$ and the fact that some electrons
start to be produced when the second laser reaches the $N$ level
of tuning are {\underline{objective}} facts,
on which of course both observers agree.
The only difference in the way in which the measurement procedure
is perceived by the two observers is the calibration.
From the point of view of $O'$
the threshold is reached in a situation that corresponds
to having one laser tuned at energy $\epsilon' = n {\cdot} \Delta E'_a$
and the other laser at energy $E' = N {\cdot} \Delta E'_b$.

Lorentz symmetry governs varius aspects of this experimental
setup. Specifically, for known $V_0$,
Lorentz symmetry governs the relation between $\epsilon$,$E$
and $\epsilon'$,$E'$ (but this part is not specific
to threshold experiments); moreover, Lorentz symmetry predicts that,
independently of  $V_0$, the product $\epsilon {\cdot} E$ will have
the same value as the product $\epsilon' {\cdot} E'$
and these products will give the square of the electron
mass: $\epsilon {\cdot} E = \epsilon' {\cdot} E' = m_e^2$.

The threshold condition $\epsilon {\cdot} E = m_e^2$
is of course the comparison of two relativistic invariants:
the invariant $m_e^2$ of the emerging electron-positron pair
and the invariant $\epsilon {\cdot} E$ of the system of two photon
colliding head on.
The fact that the production of an electron-positron pair
is an objective fact that can be witnessed by two (or more)
observers imposes that the threshold condition be
a fully invariant statement\footnote{In principle
an invariant statement does not need to be based on
one of the relativistic invariants of the relativistic theory.
For example a logically consistent threshold condition
for the process $\gamma + \gamma \rightarrow e^+ + e^-$
can be formulated by making reference to center-of-mass frame.
The condition could state that in the center-of-mass frame
the kinematics of the process must enjoy a certain property.
An observer for whom the center of mass of the process is not
at rest would then have to first boost the observed
energies and momenta to center of mass frame and then apply the
condition. In ordinary special relativity one could say that
the condition for energy-momentum conservation is to be imposed
in the center-of-mass frame, but, since the ordinary Lorentz transformations
preserve the energy-momentum-conservation conditions, this authomatically
implies that energy-momentum conservation is satisfied in every frame.
In some alternative relativistic theory it would be logically consistent
to introduce a threshold condition in the center-of-mass frame which
is not preserved by the transformation rules.}: when satisfied for one
inertial observer it must satisfied also for all other
inertial observers. This is of course true in Special Relativity,
as a result of the properties here reviewed of Lorentz symmetry.
However, this point has wider validity:
the threshold condition
must be an invariant statement in all physical theories
of particle collisions, since it is unacceptable that
some observers would see two photons disappearing in
an electron-positron pair while others would see the
two photons crossing each other without particle production.

\section{Comparison of space-rotation symmetry
and Lorentz symmetry}
In the previous two sections some aspects of
space-rotation symmetry
and Lorentz symmetry have been revisited.
My emphasis has been on the way in which these symmetries
characterize the relations between certain experimental results.
Some analogies have emerged, which I want to summarize here
using as examples the angular momentum vector $\vec{L}$
and its three components ($L_x,L_y,L_z$), for what concerns
space-rotation symmetry, and the world-sheet ${\cal W}$
and its associated properties, equal-time area $A$ and surface
velocity $V$, for what concerns Lorentz symmetry.
I have observed that in order to predict the value of $L_{x'}$,
the component of the objective entity $\vec{L}$
along a certain direction
(possibly identified with a magnetic field that specifies
the ``$x$ axis" of a second observer $O'$)
the observer $O$ must do a triple measurement, a measurement
of $L_x$, $L_y$  and $L_z$. The value of $L_{x'}$
cannot be predicted if $O$ only measures $L_x$.
This was a key conceptual ingredient for my analysis
of the compatibility between angular-momentum discretization
and continuous space-rotation symmetry.

Analogously
Lorentz symmetry makes definite predictions for $A'$
(the equal-time surface area according to observer $O'$)
when $V$ and $A$ (the surface velocity and the equal-time surface area
according to observer $O$) are known.
If instead only $A$ is known nothing can be said about $A'$.
Discretization of areas can therefore be compatible
with Lorentz symmetry if $V$ is appropriately undertermined
on $A$ eigenstates (just like $L_x$ discretization in
ordinary quantum mechanics is accompanied by
$L_x$,$L_y$ noncommutativity such that $L_y$ is undetermined
on $L_x$ eigenstates in a way  appropriate for the preservation
of space-rotation symmetry).

However, the analogy between the analysis of the interplay
between Lorentz symmetry and area discretization
and the analysis of the interplay between space-rotation
symmetry and angular-momentum discretization must not
be pushed too far.
On the Lorentz-symmetry side from the
world-sheet ${\cal W}$ an observer $O$ with a specific
choice of time axis (a clock) obtains both the
observable ``velocity of the
surface" $V$ and ``area of the surface" $A$.
On the space-rotation-symmetry side from the
angular-momentum vector $\vec{L}$ an observer $O$ with
a specific choice of $x$ axis (a magnetic field) obtains
only the observable $L_x$.
Noncommutativity of $L_x$ with $L_{x'}$ looks more like
the noncommutativity of $A$ and $A'$, rather than
the noncommutativity of $V$ and $A$.
In fact, $L_x$ and $L_{x'}$ are projections of the same
objective entity, angular momentum vector, $\vec{L}$
along two (space) axes, just like $A$ and $A'$
are ``projections" of the objective entity, world-sheet, ${\cal W}$
that are specified by two choices of (time) axis.
However, given ${\cal W}$, a single choice
of time axis allows to specify both $V$ and $A$; moreover,
the knowledge of $V$ and $A$ for one choice of time axis
allows to predict the values of $A'$ (and $V'$)
for other choices of time axis. Instead if only one
(space) axis is introduced the projection of $\vec{L}$
along that axis only gives us one observable, $L_x$,
and the knowledge is not sufficient to predict the value
of $L_{x'}$. Of course, I am using the same term ``projection"
to describe what are very different operations on the
angular-momentum vector and on the world-sheet.
This however might have important implications about
how a change of ``projection axis" should affect the
observables that are significant from the perspective
of the symmetries.

Having failed to achieve any deeper understanding of the
possible role of this (possibly even insignificant) point,
I am however tempted to conjecture that it might be related
to some of the observations which, on the measurement-analysis
side I reported in the previous section.
In Subsections 3.3, 3.4 and 3.5 I observed that some measurement
procedures that are relevant for Lorentz symmetry can be
shared (simultaneously witnessed) by different observers.
For example, the measurement procedure in Subsection~3.4
allows (at once) that both observer  $O$ and observer $O'$
measure the distance  $\overline{AC}$, finding results $L$
and $L'$ respectively. As the careful reader can easily
verify, the same argument applies to area measurement
(although the analysis of the measurement procedure is
somewhat more complex). Instead for the measurement
of components of angular momentum I was unable
to find a similar situation: I couldn't find a context
in which a single measurement procedure intended for
the measurement of $L_x$ ended up also giving a measurement
of $L_{x'}$.
This might related with the fact that the measurement
procedures for lengths,
areas (and similar)
usually (necessarily?) assume the knowledge of the velocity
of the segment (surface) whose length (area) is being measured.
These observations, however tentative, appear to be potentially
relevant for developing an intuition for what
to expect of a quantum-spacetime theory on the subject of
lengths/areas measurement and Lorentz symmetry.

\section{Various scenarios for the fate of Lorentz symmetry
in quantum spacetime}
In order to proceed in the spirit of my analysis of
space rotations in ordinary quantum mechanics
we need to identify from the previous Section some Lorentz-symmetry
characteristic measurements, measurements for which Lorentz symmetry
governs the relations between the numerical values obtained
by different observers. For the case of space-rotation symmetry
my discussion focused on the measurement ($L_x,L_y,L_z$), simultaneous
measurement of the three components of the angular momentum of
a system, the measurement of $L_x$ only, and the measurement
of $L^2$. The measurement of ($L_x,L_y,L_z$) is relevant for classical
space-rotation symmetry, through the associated prediction
of ($L_{x'},L_{y'},L_{z'}$),
but it is not an allowed measurement
in quantum mechanics (not in the classical sense,
which would require simultaneous eigenstates of all three operators),
so at the quantum level the measurement of ($L_x,L_y,L_z$)
cannot be used to test space-rotation symmetry. However, one
can test the validity of the space-rotation transformation
rules on expectation values of ($L_x,L_y,L_z$), and in ordinary
quantum mechanics this test is successful (the symmetry does hold).
The measurement of $L_x$
is allowed both at the classical and at the quantum level,
but space-rotation symmetry makes no prediction concerning this
measurement: the knowledge of $L_x$ does not allow to reconstruct $L_{x'}$
($x'$ being the $x$ axis of another, rotated, observer).
The measurement of $L_x$
is not meaningful for Lorentz-symmetry transformations:
Lorentz symmetry does not govern
the relation between the numerical values of $L_{x}$ and $L_{x'}$.
The measurement of $L^2$ is allowed both at the classical
and at the quantum level,
and space-rotation symmetry describes it as an invariant
(no effect of the classical continuous symmetry on
the discretization).

Let us consider a few measurements that are relevant on the
Lorentz-symmetry side. I start with the observables length, $L$,
area, $A$, volume, $\Omega$ and time, $\tau$.
If the observer $O$ measures only the length of a bar (but not its
velocity) Lorentz symmetry makes no prediction on the value
that another observer $O'$ would attribute to that length.
Similarly for the area of a surface and the volume of an object.
And, again similarly, if the observer $O$ measures only the value $\tau$
of the ticks of a light clock (but not the velocity of the light clock)
Lorentz symmetry makes no prediction
for the value $\tau'$ of the
corresponding measurement done by $O'$.
If the observer $O$ measures both the length of the bar and its
velocity, a $(V,L)$ measurement,
then Lorentz symmetry makes a definite prediction.
Lorentz-symmetry (active) transformations
dictate that actually\footnote{For simplicity, but without any
true loss of generality, I am basically discussing a one-dimensional
configuration: the bar's end points both are on the $x$ axis of $O$
and the relative $OO'$ velocity is also along that $x$ axis.}
$(V,L)=(V,\sqrt{1-V^2/c^2}L_0)$, where $L_0$
is the rest length of the bar.
For given relative $OO'$ velocity, $V_0$, Lorentz symmetry
predicts the velocity $V'$ of the bar with respect to $O'$,
and predicts that $L'=\sqrt{1-V'^2/c^2}L_0=\sqrt{(c^2-V'^2)/(c^2-V^2)}L$.
So Lorentz symmetry is fully operative on the $(V,L)$ measurement:
it predicts the transformation $(V,L) \rightarrow (V',L')$.
Completely analogous remarks apply
to the measurements $(V,A)$, $(V,\Omega)$ and $(V,\tau)$.

The measurements $(V,L)$, $(V,A)$, $(V,\Omega)$, $(V,\tau)$
are relevant for Lorentz symmetry just like the measurement
($L_x,L_y,L_z$) is relevant for space-rotation symmetry: the classical symmetry
makes definite predictions for the laws of transformation of
these measurements and they are not invariants (the numerical values
of the observables do change between inequivalent classes of
observers).
An example of Lorentz-symmetry invariant is of course $E^2-c^2 \vec{p}^2$
($E$,$p$ energy-momentum of a particle).
The measurement of $E^2- c^2 \vec{p}^2$
is relevant for Lorentz symmetry just like the measurement
of $L^2 \equiv L_x^2 + L_y^2 + L_z^2$
is relevant for space-rotation symmetry: the classical symmetry
makes definite predictions that
these measurements correspond to invariants
(the numerical values of the observables is the same for all
inertial observers).

The understanding of the fate of Lorentz symmetry in quantum spacetime
requires us to establish if, and in which way, the
spacetime quantization affects these physical predictions
for non-invariant Lorentz-symmetry-meaningful measurements such
as $(V,L)$, $(V,A)$, $(V,\Omega)$, $(V,\tau)$
and for invariant Lorentz-symmetry-meaningful measurements such
as $E^2- c^2 \vec{p}^2$.
Even before considering specific quantum-gravity proposals
it is possible to discuss in general terms a few scenarios
for the fate of Lorentz symmetry in quantum gravity.

\subsection{Classical Lorentz symmetry preserved}
Of course, it is plausible that classical Lorentz symmetry
applies to the (still to be established)
correct theory of quantum gravity in the same sense that
space-rotation symmetry applies in ordinary quantum mechanics.
This is actually the most natural expectation for quantum-gravity
theories in which spacetime is not really quantized, in the
sense that these theories still rely on a classical background
spacetime and admit classical Minkowski spacetime as a
possible background. An example of quantum-gravity theory which
reflects this expectation is string theory.
In fact, in string theory among the admitted spacetime
backgrounds it is still possible to choose spacetimes that are
completely classical. In that case physical processes still occur
in a classical (background) spacetime arena, and
spacetime is only ``quantized"
in the sense that some new particles (notably, the graviton)
are allowed to propagate in this fundamentally classical spacetime
and mediate gravitational interactions.
Of course, departures from Lorentz symmetry are not necessary
as long as a theory (as in the case of string theory) still admits
the possibility of a background spacetime that is exactly Minkowski.

In other quantum-gravity approaches classical Minkowski spacetime
will only emerge as an approximate description of a fundamentally
nonclassical spacetime; for example, certain noncommutative versions
of Minkowski spacetime are perceived just like classical Minkowski
spacetime by long-wavelength particles, while they are fully nonclassical
for short-wavelength particles. In those cases Lorentz symmetry
will only be an approximate symmetry, emerging in the low-energy
limit.
At present this is not the case in string theory, but it is
noteworthy that the issue of the ``spacetime background" is probably
the one on which most progress must be sought within the string-theory
research programme. The idea of a background-dependent approach
to the quantum-gravity problem is not satisfactory on many grounds.
Future developments in the string-theory programme might eliminate
the need to make reference to a background spacetime, and at that stage
it might be interesting to reassess the status of exactly classical
Minkowski spacetime within string theory.

\subsection{Classical Lorentz symmetry not preserved}
The situation is quite different in theories that really change the
fundamental description of spacetime, such as theories that invoke
some form of spacetime discretization or spacetime noncommutativity.
In these instances it is actually more natural to expect that
the fate of Lorentz symmetry be nontrivial, that classical Lorentz
symmetry would not apply to such quantum-gravity theories,
at least not in the strong sense in which classical space-rotation
symmetry applies to ordinary quantum-mechanics theories.
The point is that classical space-rotation symmetry ``survived"
(in the sense clarified in Section~2) the advent of ordinary
quantum mechanics because space-rotation symmetry pertains to spacetime
and ordinary quantum mechanics still relies on a fully classical
spacetime. The idea of spacetime quantization would instead truly
modify the structure of spacetime and it is therefore natural to expect
(as verified in certain specific toy-model examples)
that the spacetime symmetry we call Lorentz symmetry
would be affected by the spacetime quantization.
In the next Section I discuss some examples of
noncommutative versions of Minkowski spacetime and clarify
that departures from ordinary Lorentz symmetry are rather natural.
Also the idea of spacetime discretization naturally leads to
the expectation of a nontrivial fate for the continuous
classical Lorentz symmetry, but our intuition is still
not reliable in these contexts.
Clearly a simple-minded rigid discretization of Minkowski spacetime
would not be consistent with the
continuous classical Lorentz symmetry~\cite{thooftdiscrete},
but it is difficult to develop some intuition
for more sophisticated ways to introduce discreteness in
spacetime structure. An interesting attempt to introduce
a non-trivial discretization of spacetime structure has emerged
from the ``loop quantum gravity"~\cite{carloliving,lqga,lqgb,lqgc}
research programme. Section~7 is devoted to this loop-quantum-gravity
discretization scenario and there I shall argue that the present
understanding/interpretation of certain loop-quantum-gravity
results appears to be in conflict with classical Lorentz symmetry,
but I shall also argue that at present it is not clear whether
this is a genuine feature of the theory or perhaps just an indication
that the relevant results should be interpreted and analyzed
more carefully.

\subsection{Deformation of the classical Lorentz symmetry}
In the preceding Subsection I made the point that in approaches
that rely on a genuinely nonclassical spacetime it is natural
(though perhaps not necessary) to find that the fate of
classical Lorentz symmetry is nontrivial, {\it i.e.}
that classical Lorentz symmetry would not be an exact
classical symmetry at the quantum-spacetime level.
The present Subsection comments on a scenario for the fate of Lorentz
symmetry in quantum spacetime which this author
finds rather appealing: the scenario in which one basically still
has the same conceptual structure of Lorentz symmetry
(for example with six symmetry generators) but Lorentz
transformations are deformed in such a way as to have two
observer-independent scales,
a length scale $\lambda$ (possibly given by the Planck length)
and (again, as in ordinary Lorentz symmetry)
the speed-of-light constant.

This deformed-symmetry scenario, which is being called ``Doubly
Special Relativity",
was proposed by this author in Ref.~\cite{dsr1dsr2},
where a first example of such deformed symmetries
was also constructed..
A second example was more recently constructed by
Maguejio and Smolin~\cite{leedsr},
and even more recently Kowalski-Glikman and Nowak~\cite{jurekdsrNEW}
and Lukierski and Nowicki~\cite{lukiedsr}
have reported progress in the construction of
a larger class of such deformed symmetries.

It appears that the idea of deformed Lorentz symmetry
is indeed realized in certain nonclassical pictures
of spacetime, at least in certain noncommutative
versions of Minkowski spacetime. I shall comment
on this in greater detail in the next Section.
Here I want to give a physical characterization of
a deformation of Lorentz symmetry, in the spirit I have
adopted throughout this paper.
A deformation of Lorentz symmetry would still be characterized
by transformation rules and invariants of the transformation
rules, but their structure would be different from
the one of ordinary Lorentz symmetry.
For example, just like Lorentz symmetry predicts that
$E^2- c^2 \vec{p}^2$ is an energy-momentum invariant,
the deformation of Lorentz symmetry described in
Ref.~\cite{dsr1dsr2} predicts that\footnote{Focusing on this
specific form of the dispersion relation, while not necessary,
can be motivated by previous arguments in the study of
quantum algebras~\cite{majrue,kpoinap} and of
an approach to the study of noncritical
string theory~\cite{aemn1}. The quantum-algebra results
are found~\cite{dsr1dsr2} (at least for
the one-particle sector~\cite{dsr1dsr2})
to provide an acceptable description of
DSR transformations (doubly-special-relativity transformations),
in the same sense that the preexisting Lorentz transformations
provided the correct mathematical language for Einstein's
Special Relativity. DSR transformations can accordingly
be called quantum algebra or ``$\kappa$-deformed"~\cite{majrue,kpoinap}
transformations, just like we refer to the special-relativistic
transformation rules of Einstein's physical theory as Lorentz
transformations.}
$\lambda^{-2} c^{-2} (e^{\lambda E/c}
+ e^{-\lambda E/c} - 2) - c^2 \vec{p}^2 e^{\lambda E/c}$
is an invariant. In ordinary Lorentz symmetry one has
the observer-independent scale $c$ that plays the role
of speed of massless particles of any energy
and maximum attainable velocity,
whereas in the deformation of Lorentz symmetry described in
Ref.~\cite{dsr1dsr2} one has
the observer-independent scale $c$ that plays the role
of speed of low-energy massless particles,
and the observer-independent scale $\lambda$ that plays the role
of inverse of the maximum attainable momentum.
These are physical characteristics of the deformed symmetry
that could be tested experimentally
and one can also attempt to construct theories whose predictions are
consistent with these characteristics of the symmetry.

Notice that a deformation does not involve any loss
of symmetry and does not involve any changes in the
type of rules that we associate to the
concept of symmetry, {\it i.e.} a deformation of Lorentz symmetry
can still be a classical symmetry according to the definitions
here given. This idea might find application in quantum pictures of
spacetime that enjoy classical symmetries:
in this scenario the transition from
classical to quantum spacetime would require that the symmetries
(still being classical in nature) reflect/preserve the new
(quantum) structure introduced in spacetime by quantization.
An example of this scenario will be discussed in Subsection~6.1.

\subsection{Lost Lorentz-symmetry}
The idea of a deformation of Lorentz symmetry, considered
in the previous Subsection, does not involve any loss
of symmetry, but it is of course legitimate
(although rather painful conceptually) to consider the
possibility that the transition from classical to quantum spacetime
would actually cause a loss of symmetry.

Let me provide a physically-characterized example of
loss of symmetry by considering again the dispersion relation.
In the previous Subsection I implicitly considered two
dispersion relations, the standard $E^2- c^2 \vec{p}^2 = c^4 m^2$
and the dispersion relation $\lambda^{-2} c^{-2} (e^{\lambda E/c}
+ e^{-\lambda E/c} - 2) - c^2 \vec{p}^2 e^{\lambda E/c} = c^4 m^2$
characteristic of a deformed symmetry scenario (if the laws
of transformations between inertial observers are accordingly
deformed~\cite{dsr1dsr2}).
An example of dispersion relation that would signal symmetry loss
is $E^2- c^2 \vec{p}^2 + E u_0-  \vec{p} {\cdot} \vec{u} = c^4 m^2$,
with $u$ a given four vector that transforms from observer to
observer according to Lorentz transformations,
so that the values of $u_0$, $u_x$, $u_y$ and $u_z$
are not identical to the values of $u_0'$, $u_x'$, $u_y'$ and $u_z'$.
Of course, the presence of $u$ allows one to identify a
preferred class of inertial observers (specified by a chosen
value of the components of $u$ along the $t$,$x$,$y$,$z$ axis
of those observers), signaling a loss of Lorentz symmetry.

A key point here is that, according to my definitions,
a genuine loss of Lorentz symmetry will be present only
if an object such as the $u$ of my example is an intrinsic
property of spacetime. We already know (even experimentally)
that in a perfectly Lorentz invariant theory living in a
perfectly Lorentz invariant spacetime one can have cases
in which the dispersion relation involves some external
four-vector or tensor. For example, we know that a Lorentz-invariant
theory of the propagation of light in certain media does
predict a modification of the dispersion relation, often even involving
preferred directions (the preferred directions of the
dispersion relation reflect the preferred directions of the medium).
For the case of light propagating in a medium we can still (and should)
speak of a Lorentz-symmetric theory living in a Lorentz-symmetric
spacetime in which the specific system under study (in particular
the medium) is not invariant under Lorentz transformations.
It is clearly a different situation when spacetime itself
has preferred directions or anyway allows the identification
of a preferred class of inertial observers. In such cases
it is appropriate to speak of loss of Lorentz symmetry.

\subsection{Classical Lorentz symmetry spontaneously broken}
An important class of scenarios in which there is loss of Lorentz
symmetry is the one in which Lorentz symmetry is spontaneously broken.
Here of course I have in mind the field-theory/particle-physics
mechanism of spontaneous symmetry breaking.
It is not easy to imagine a similar mechanism applied
to spacetime structure, especially because we lack a true understanding
of the concept of spacetime vacuum (we can perhaps attempt to
describe the concept of empty spacetime, but it is much harder
to imagine some sort of minimum-energy spacetime,
since we do not even have an {\it a priori}
concept of energy for contexts in which a background spacetime
is not provided {\it ab initio}).
However, it is not unplausible that the correct theory of
spacetime physics would enjoy Lorentz symmetry in the sense
that different spacetime solutions that are connected
by a Lorentz transformation are equally likely, but then
the most likely solutions (``the vacuum") would not themselves
enjoy Lorentz symmetry. (This of course requires that the appropriate
concept of ``most likely spacetime solution" does not identify
a single spacetime solution, but rather identifies a 6-parameter
family of degenerate solutions, all mapped one into the other
by Lorentz transformations.)
The example of a four-vector $u$ discussed in the previous
Subsection could emerge in such a spotaneous-symmetry-breaking scenario:
the ``vacuum solutions" would be characterized by $u$, and
all forms of $u$ that are connected by a Lorentz transformation
to a certain $u^*$ would all be equally likely, but then
Nature would have chosen a specific vacuum, breaking the degeneracy.

\subsection{Fuzzy Lorentz symmetry?}
My {\it a priori} discussion of possible scenarios for
the fate of Lorentz symmetry in quantum spacetime
cannot aim for completeness. Nature may well host
a scenario which this author has not managed
to even imagine. A characteristic of all scenarios
I have considered up to this point is that they
still rely on the concept of a classical symmetry (in the
sense introduced in this paper). The classical symmetry is
deformed or even violated (symmetry loss) but the questions
one would ask (the properties that characterize the symmetry concept)
are formulated classically in the sense of Section~2.
To this author it is not even clear whether one should/could contemplate
anything different from this.
One is confronted with similar conceptual challenges when trying
to analyze the conceptual framework of ordinary quantum mechanics
without relying on a classical apparatus (what is a nonclassical
apparatus? what would be a nonclassical interpretation of
the readout of a measurement device?).
At least for the context of
passive symmetry transformations, which is more constrained
by demands of the objectivity of physical processes,
it is difficult to think about alternatives to the
conceptual framework of classical symmetries.
Think for example of the description of the length-contraction
experiment on which I focused in Subsection 3.3.
There the two observers not only measure the same length
but they also rely on a single measurement procedure
(which however they interpret in a different way).
The two observers even agree on the
readout that gives the result of that length measurement
experiment (which is the number $N$ of ticks of the light clock),
and they obtain a different result for the length measurement
simply because their relative motion affects the calibration
attributed to the devices and the description of the measurement
procedure. The (passive) symmetry transformation connects
the interpretation of $N$ for one observer with the interpretation
of $N$ for another observer, and it is difficult to
imagine that the relation between these interpretations would
not be classical in the sense here advocated.
It might only be possible in theories predicting some new limitations
(of an appropriate type) on the mechanism of calibration of devices
(or perhaps an absolute limitation on the measurability of
the relative velocity of two observers).

Still, it is tempting to conjecture here that it might be
possible for a quantum theory of spacetime to accomodate
some sort of nonclassical, ``fuzzy", symmetry concept.
Perhaps a symmetry concept which only applies to ensembles of
observations and not to a single observation. However, this possibility
is indeed challenged\footnote{Since the same length-measurement
procedure can be shared by two observers it would
be paradoxical if, for example, one observer was to conclude,
after repeated measurements, that she is dealing with a length
eigenstate, while the other observer would conclude that he is dealing
with a superposition of different eigenstates.
Both observers see the same readouts $N$ of the light clock
and they must therefore agree on whether or not they are dealing
with a length eigenstate.}
by the analysis of contexts,
such as the length measurement in Subsection 3.3,
in which a single measurement is meaningful for
two observers and therefore the symmetry transformation should
predict how that single measurement procedure appears
to the two observers.

\subsection{Challenge to quantum-gravity theories}
In this Section devoted to a brief description of various
scenarios for the fate of Lorentz symmetry in quantum spacetime
it appears to be appropriate to formulate a challenge to
quantum-gravity theories. A large research effort is focusing
on the quantum-gravity problem, but only a relatively small
percentage of these studies concerns the fate of Lorentz
symmetry. The recent progress (and the expected progress)
of sensitivity of tests of Lorentz symmetry (see, {\it e.g.},
Refs.~\cite{grbgac,gampul,billetal,ita,gactp2,glast}) renders this
state of affairs unjustifiable.
If any deviation from ordinary (classical-spacetime) Lorentz
symmetry is hosted by a quantum-gravity theory, it will
most likely turn out that the associated
predictions can be tested with very high accuracy.

Since the ultimate goal is comparison to experimental
results, the fate of Lorentz symmetry must be analyzed in
quantum-gravity theories giving priority to physical issues
(the nature and magnitude of the predicted effects)
rather than formalism issues.
It is in this respect that there is a clear set of challenges
to quantum-gravity theories.
For example, we are reaching extremely high sensitivity~\cite{grbgac}
to the study of the propagation over cosmological distances
of short-duration bursts of photons.
In classical physics the distance travelled, $L$,
would be classical, the photons would be point-like
and the photons would follow the classical trajectory along $L$,
so, according to classical physics, a
group of such photons
which were emitted
simultaneously at time $t=0$
would complete the journey simultaneously at time $t=L/c \equiv T$.
Ordinary (known) quantum properties of matter (in classical spacetime)
already modify this picture: the structure of quantum mechanics
imposes that the time of emission of a particle with energy uncertainty
$\delta E$ can
only be specified with accuracy $1/\delta E$, and there is of course
a corresponding limitation on how accurately the simultaneity of
the times of arrival can be established, but the relation $T=L/c$
will emerge if appropriate averaging over a large number of
observations is performed.
The quantum properties of the particles
(still assuming classicality of the spacetime)
introduce a nonsystematic effect, an
uncertainty: $T = L/c {\pm} \delta T_{QM}$.
I have here a clear opportunity for a {\bf well-defined challenge
to quantum-gravity theories: how does a given quantum-gravity
theory affect this prediction?}
In order to be covered on all possible fronts we should be open
to the possibility of both systematic and nonsystematic quantum-gravity
effects. This can be captured in the formula
\begin{equation}
T = (L/c + \Delta T_{QG}) {\pm} \delta T_{QM} {\pm} \delta T_{QG} ~,
\label{generalTL}
\end{equation}
with self-explanatory notation.
A nonvanishing prediction for $\Delta T_{QG}$ would require a departure
from classical Lorentz symmetry.
$\Delta T_{QG}$, if nonzero, would likely be energy-dependent
(at low energies we have good data that strongly support $\Delta T_{QG}=0$)
and this would affect
the propagation over cosmological distances
of short-duration bursts of photons in an obvious way:
one would expect a systematic energy-dependent time-of-arrival
difference in the analysis of the short duration bursts
in different energy channels of our detectors.
The possibility that $\Delta T_{QG}=0$
but the given quantum-gravity theory predicts
a nonvanishing value for $\delta T_{QG}$
does not necessarily require a deviation from classical
ordinary Lorentz symmetry. The careful reader will easily realize that,
with the definition here adopted of a classical symmetry in a
nonclassical theory, it is necessary to analyze the properties of
a given $\delta T_{QG}$ picture in order to establish whether or not
it implies a deviation from classical ordinary Lorentz symmetry.
It might well be that $\delta T_{QG} \ne 0$ in a way that still
satisfies the conditions for
classical ordinary Lorentz symmetry in the given quantum-gravity theory.
A $\delta T_{QG} \ne 0$ is to be expected with rather natural assumptions
about quantum gravity: just like the quantum properties of matter
(the relevant photons) in classical spacetime introduce a nonvanishing
$\delta T_{QM}$, the quantum properties of spacetime should introduce
a nonvanishing $\delta T_{QG}$.
The magnitude and structure ({\it e.g.} the dependence on the
energy of the particles involved)
of $\delta T_{QG}$ will vary strongly from one quantum-gravity
theory to another.
A nonvanishing $\delta T_{QG}$ would of course affect
the propagation over cosmological distances
of short-duration bursts of photons.
A nonvanishing energy-dependent
value of $\delta T_{QG}$ could be established by
searching for differences in the time spread of the burst
in different energy channels
of our detectors.
A nonvanishing distance-travelled-dependent
value of $\delta T_{QG}$ could be established by
searching for differences in the time spread of bursts
with otherwise similar characteristics but reaching us
from different distances.
If $\delta T_{QG}$
is distance-independent and energy-independent
it might be hard to find experimental evidence for it
(in that case a natural estimate for $\delta T_{QG}$
would be $\delta T_{QG} \sim t_p$, and the Planck time $t_p$
is so small that
we would never find evidence for such a $\delta T_{QG}$).

The study of the propagation over cosmological distances
of short-duration bursts of photons
is clearly a key challenge to quantum-gravity theories.
Another example of important challenge for quantum-gravity
theories comes from the analysis of the energy
thresholds for certain particle-production processes,
such as the electron-positron pair production in photon-photon
collisions, which I considered in the preceding Section.
Classically two photons can have sharply-defined energies,
say $E$ and $\epsilon$, and the
process $\gamma + \gamma \rightarrow e^+ + e^-$
is allowed when $E \epsilon \ge m_e^2$ and it is absolutely
forbidden if $E \epsilon < m_e^2$.
Within ordinary quantum mechanics (here the relevant
formalism is field theory in Minkowski spacetime)
one would most naturally consider photons prepared with
energies $E$ and $\epsilon$, with $E_0 - \Delta < E < E_0 + \Delta$
and $\epsilon_0 - \delta < \epsilon < \epsilon_0 + \delta$,
but one still has a definite prediction: the process is allowed
if $E_0 \epsilon_0 \ge m_e^2 - \Delta \epsilon_0
- \delta E_0$
while it is forbidden
if $E_0 \epsilon_0 < m_e^2- \Delta \epsilon_0
- \delta E_0$.
Here there is another clear opportunity for a {\bf well-defined challenge
to quantum-gravity theories: how does a given quantum-gravity
theory affect this prediction?}
Assume we prepare the photons just as we do now,
with $E_0 - \Delta < E < E_0 + \Delta$
and $\epsilon_0 - \delta < \epsilon < \epsilon_0 + \delta$.
Will the condition
$E_0 \epsilon_0 \ge m_e^2 - \Delta \epsilon_0 - \delta E_0$
still hold?
or will it take the form
$E_0 \epsilon_0 \ge m_e^2 + \Delta_{threshold,QG}
- \Delta \epsilon_0 - \delta E_0$ ?
A nonvanishing value of $\Delta_{threshold,QG}$
would be predicted~\cite{dsr1dsr2}
in most quantum pictures of spacetime whose symmetries
are not described by ordinary classical Lorentz symmetry
(in these spacetimes even pair production by
classical photons, with ideally sharp
definition of their energies, would be governed by
$E_0 \epsilon_0 \ge m_e^2 + \Delta_{threshold,QG}$
rather than $E_0 \epsilon_0 \ge m_e^2$).
Even quantum-gravity theories in which ordinary classical
Lorentz symmetry holds (in the sense here advocated)
might predict a threshold of the form
$E_0 \epsilon_0 \ge m_e^2 + \Delta_{threshold,QG}
- \Delta \epsilon_0 - \delta E_0$,
but there the threshold deformation should be attributed to
some sort of new uncertainty principle~\cite{uncpap}.
In both cases it might not be hard to obtain stringent limits
on the predicted deviations from the classical-spacetime threshold;
in fact, the analysis~\cite{crdata}
of ultra-high-energy cosmic rays
and other observations that are primarily of interest
in astrophysics can be used~\cite{ita,gactp2,aus}
to test the idea of threshold deformation with
extremely high sensitivity.

\section{On the fate of Lorentz symmetry in noncommutative spacetime}
Noncommutative geometry is being used more and more extensively in attempts
to unify general relativity and quantum mechanics. Some quantum-gravity
approaches explore the possibility that noncommutative geometry might
provide the correct fundamental description of spacetime, while in other
approaches noncommutative geometry turns out to play a role at the level of
the effective theories that describe certain aspects of quantum gravity.

For the issues here under consideration it is the case of noncommutative
versions of flat (Minkowski) spacetime that is of interest.
Of course, a flat noncommutative spacetime could not possibly
provide a full solution to the quantum-gravity problem,
but if it turns out to be true that noncommutative geometry
is the correct language/formalism for the
description of the fundamental structure of spacetime
then in particular the spacetimes that we perceive as
(approximately) flat and classical should also be described
by noncommutative geometry and noncommutative versions of Minkowski
spacetime might therefore be relevant.

Two simple examples~\cite{starwess} are ``canonical noncommutative
spacetimes" ($\mu,\nu,\beta = 0,1,2,3$)
\begin{equation}
\left[x_\mu,x_\nu\right] = i \theta_{\mu \nu}
\label{canodef}
\end{equation}
and ``Lie-algebra noncommutative spacetimes"
\begin{equation}
\left[x_\mu,x_\nu\right] = i C^\beta_{\mu \nu} x_\beta ~.
\label{liedef}
\end{equation}
These two simple examples of noncommutative spacetimes\footnote{In this
study I will only be concerned with the simple case in which
noncommutative geometry takes the form of noncommuting coordinates,
whose commutators are either constant or depend on the coordinate
themselves. This is mostly in the spirit of an approach
to noncommutative geometry that originates
from the quantum-group research
programme~\cite{majidbook}.
Another interesting case of noncommutative version
of Minkowski spacetime, which is not consider here (but will be analyzed
in a forthcoming publication~\cite{gacjureksnyder}),
is the Snyder spacetime~\cite{snyder,jurekdsrNEWtwo},
in which however the commutators of the coordinates
are expressed in terms of elements of the Lorentz algebra.
Noncommutative
geometry is also being developed following another
approach that
originates primarily from original work by Connes~\cite{connesbook},
but in that approach nothing significant as emerged
concerning noncommutative versions of Minkowski spacetime.}
are also useful for illustrative purposes, since they provide
well-defined models in which some of the scenarios for the fate of
Lorentz symmetry in quantum spacetime considered in the previous Section
are realized.

On the  Lie-algebra side I will focus for simplicity on the
$\kappa$-Minkowski~\cite{majrue,kpoinap} spacetime ($l,m = 1,2,3$)
\begin{equation}
\left[x_m,t\right] = {i \over \kappa} x_m ~,~~~~\left[x_m, x_l\right] = 0 ~,
\label{kmindef}
\end{equation}
which is one of the most studied\footnote{$\kappa$-Minkowski spacetime
is a Lie-algebra spacetime that clearly enjoys classical space-rotation
symmetry (while boosts are deformed). I think there could be justifiable
interest in the possibility of studies
of other flat space-rotation-invariant
spacetimes based on the Lie-algebra-type
algebraic relations, such as  $[x_i,x_j]= i {\cal L} \epsilon^{ijk} x_k$
which would also naturally lead to a theory with classical space rotations
and deformed boosts. However, at least within the presently-adopted
mathematical framework for these noncommutative geometries,
from
the algebraic relations $[x_i,x_j]= i {\cal L} \epsilon^{ijk} x_k$
one is naturally led~\cite{fuzzysphere} to the description of
spheres rather than flat spacetimes.}
noncommutative to classical Minkowski spacetime.

\subsection{$\kappa$-Minkowski noncommutative spacetime}
Detailed analyses of $\kappa$-Minkowski spacetime can be found
in Refs.~\cite{majrue,kpoinap,gacmaj}. Here I just want to
provide an intuitive characterization of the fate of Lorentz
symmetry in this noncommutative version of Minkowski spacetime.
A first point is that the law of composition of momenta is
deformed and nonlinear in $\kappa$-Minkowski. This is encoded
in the so-called coproduct.
An intuitive way to see this is through the introduction
of the Fourier tranform.
It turns out~\cite{gacmaj,majoek,lukiestar} that in the
$\kappa$-Minkowski case the correct formulation of the Fourier theory
requires a suitable ordering prescription\footnote{There is of course an
equally valid alternative ordering prescription in which the time-dependent
exponential is placed to the left (while we are here
choosing the convention with the time-dependent exponential to the right).}
for wave exponentials:
\begin{equation}
:e^{i k^\mu x_\mu}: \equiv e^{i k^m x_m} e^{i k^0 x_0}
~.
\label{order}
\end{equation}
These wave exponentials are actual solutions of a $\kappa$-Minkowski
wave equation~\cite{gacmaj}.
While wave exponentials of the
type $e^{i p^\mu x_\mu}$ would not combine in a simple way
(as a result of the $\kappa$-Minkowski noncommutativity relation),
for the ordered
exponential one finds
\begin{equation}
(:e^{i p^\mu x_\mu}:) (:e^{i k^\nu x_\nu}:) =
:e^{i (p \dot{+} k)^\mu x_\mu}:
\quad.
\label{expprodlie}
\end{equation}
The notation ``$\dot{+}$" here introduced reflects the
behaviour of the mentioned ``coproduct",
composition of momenta\footnote{Here I use the vague
expression ``composition of momenta". In physics we need to compose
momenta in various situations, {\it e.g.} when we combine two plane
waves into one and when we impose energy-momentum conservation
in multiparticle processes. Using the bare coproduct in the law
of composition of plane waves appears to be appropriate in light
of the property (\ref{expprodlie}), but using the bare coproduct
in the law of conservation of energy-momentum would lead to a statement
of energy-momentum conservation which is not objective for
all inertial observers~\cite{gacinprep}: for a particle-producing
collision process $a+b \rightarrow c+d$
laws of the type $(p_a \dot{+} p_b)^\mu = (p_c \dot{+} p_d)^\mu$
are inconsistent with the relevant, $\kappa$-deformed,
laws of transformation for the $p^\mu$'s of the four particles
(the condition $(p_a \dot{+} p_b)^\mu = (p_c \dot{+} p_d)^\mu$
can be imposed in a given inertial frame but it will then be violated
in other inertial frames!). This point has strangely been
missed in the whole of the $\kappa$-deformation literature
(see, {\it e.g.}, Ref.~\cite{kpoinap,lukiedsr}),
but now, in light of the recent doubly-special-relativity
proposal~\cite{dsr1dsr2}, it must be seen
as a top-priority problem for the $\kappa$-deformation
programme.} in $\kappa$-Minkowski
spacetime:
\begin{equation}
p_\mu \dot{+} k_\mu \equiv \delta_{\mu,0}(p_0+k_0) + (1-\delta_{\mu,0})
(p_\mu +e^{\lambda p_0} k_\mu) ~. \label{coprod}
\end{equation}

As argued in Refs.~\cite{dsr1dsr2} the nonlinearity of the law of composition
of momenta should require an absolute (observer-independent) momentum scale,
just like upon introducing a nonlinear law of composition of velocities
(in going from Galilei/Newton relativity to Einstein relativity)
one must introduce the absolute observer-independent scale of
velocity $c$. The inverse of the noncommutativity scale $\lambda$
plays the role of this absolute momentum scale. This of course
requires~\cite{dsr1dsr2} transformation laws for energy-momentum
between different observers which have two invariants, $c$ and $\lambda$,
while ordinary Lorentz transformations have only one invariant.
An example of laws of transformation that enjoy this property was
used as illustrative example in Refs.~\cite{dsr1dsr2}, in which
the idea of deformed Lorentz symmetry was introduced.
A key point is that the deformed Lorentz transformations form group.
They actually are a nonlinear representation of the Lorentz group itself.
While ordinary Lorentz transformations leave invariant the
combination $E^2 - c^2 p^2$,
the deformed transformation rules leave invariant the combination
\begin{equation}
C_\lambda (E,{\vec{p}}^2)=
{c^2 \over \lambda^2} (e^{\lambda E/c} + e^{-\lambda E/c} - 2)
- c^2 {\vec{p}}^2 e^{\lambda E/c}
~. \label{dispecial}
\end{equation}
The dispersion relation $E^2 = c^2 p^2 + c^4 m^2$
is accordingly replaced by the new (deformed) dispersion relation
implicitly defined by the
requirement $C_\lambda (E,{\vec{p}}^2)=C_\lambda (m,0)$.

In work that preceeded Refs.~\cite{dsr1dsr2}, some
examples of Hopf algebras that could represent
deformed {\underline{infinitesimal}} symmetry transformations
had been worked out, but it was believed~\cite{kpoinnogroup}
that these algebra structures would not be compatible with
a genuine symmetry group of finite transformations.
In Refs.~\cite{dsr1dsr2} it was proposed that one should look
for deformed transformation laws that form a genuine group
and it was shown that one example of the Hopf algebras that mathematical
physicists had been developing did allow the emergence of
a group of finite transformations (while the same is not
true for other examples of these Hopf algebras).
Interest in the proposal~\cite{dsr1dsr2} of deformed Lorentz
symmetry is growing, and very recently other examples of the same
type of deformed transformation rules have been constructed
in Refs.~\cite{leedsr,jurekdsrNEW,lukiedsr}.

\subsection{Canonical noncommutative spacetime}
Also in the case of canonical noncommutative spacetime
an intuitive characterization of the fate of Lorentz symmetry
can be obtained by looking at wave exponentials. The Fourier
theory in canonical noncommutative spacetime is based on
simple wave exponentials $e^{i p^\mu x_\mu}$ and from the
relevant noncommutativity relations one finds that
\begin{equation}
e^{i p^\mu x_\mu} e^{i k^\nu x_\nu}
= e^{-\frac{i}{2} p^\mu
\theta_{\mu \nu} k^\nu} e^{i (p+k)^\mu x_\mu} ~,
\label{expprodcano}
\end{equation}
{\it i.e.} the Fourier parameters $p_\mu$ and $k_\mu$ combine just as
usual, with the only new ingredient of the overall phase factor that depends
on $\theta_{\mu \nu}$.
The fact that momenta combine in the usual way reflects the fact that
the transformation rules for energy-momentum from one
(inertial) observer to another are still the usual, undeformed,
Lorentz transformation rules. However, the product of wave exponentials
depends on $p^\mu \theta_{\mu \nu} k^\nu$: it depends on the ``orientation"
of the energy-momentum vectors $p^\mu$ and $k^\nu$
with respect to the $\theta_{\mu \nu}$ tensor.
The $\theta_{\mu \nu}$ tensor plays the role of a background that
identifies a preferred class of inertial observers.
Different particles are affected by the presence of this background
in different ways, as shown by the results~\cite{cncft1,cncft2,cncft3}
of the study of field theories in canonical noncommutative spacetimes.

The situation, for what concerns Lorentz transformations, is actually
very familiar. We know well many contexts in which the presence
of a background selects a preferred class of inertial observers.
This is reflected, for example,
in the fact that the dispersion relation for light
travelling in water, in certain crystals, and in
other media is modified. The study of
field theories in canonical noncommutative spacetimes
shows~\cite{cncft1,cncft2,cncft3}
that the $\theta_{\mu \nu}$ background induces effects that
are somewhat similar to the ones induced by the presence of a crystal,
including the effect of birefringence of light.

The $\theta_{\mu \nu}$ tensor ``breaks" Lorentz symmetry in the same sense
that any medium breaks Lorentz symmetry: the theory is still fundamentally
Lorentz invariant but the Lorentz invariance is manifest only when
different observers take into account the different form that the
background, in this case the $\theta_{\mu \nu}$ tensor,
takes in their respective reference systems.
While the single (dimensionful) deformation parameter $\lambda$
of the $\kappa$-Minkowski
spacetime is observer-independent, {\it i.e.} takes the same value for
all observers,
the $\theta_{\mu \nu}$ matrix behaves like a Lorentz tensor:
the elements of the $\theta_{\mu \nu}$ matrix take different values
for different observers.
If the observers only take into account the transformation rules
for the energy-momentum of the particles involved in a process
the results are not the ones predicted by Lorentz symmetry;
in particular, the dispersion relation depends on the background.
In fact, the dispersion relations found
in the study of
field theories in canonical noncommutative spacetimes
acquire~\cite{cncft1,cncft2,cncft3}
a dependence on $p^\mu \theta_{\mu \nu} p^\nu$,
$E^2 = c^2 p^2 + c^4 m^2 + f(p^\mu \theta_{\mu \nu} p^\nu)$,
with the function $f$ that depends on the spin and charges of
the particle.

Concerning the construction of field theories in canonical
noncommutative spacetimes there are some interesting issues.
It has emerged that field theories constructed in strict analogy
with the way we construct them in commutative spacetimes
do not host the familiar mechanism of Wilson decoupling between
ultraviolet and infrared degrees of freedom~\cite{cncft1,cncft2,cncft3}.
This connection between ultraviolet and infrared is not necessarily
troublesome~\cite{gacluisa,gacgianluca};
moreover,
to this author it is not at all obvious that in these noncommutative
geometries one should necessarily construct field theories in
strict analogy with what usually done in commutative spacetime.
Since the $\theta_{\mu \nu}$ tensor can be used to single out
a preferred class of inertial observers one could for example
introduce a maximum momentum for that class of inertial observers,
and one could even specify the laws of physics only according to
that class of preferred observers. The other observers would of course
witness the same physical phenomena but would describe them
as Lorentz transformations of the phenomena seen by
the preferred class of inertial observers.

\subsection{Doplicher-Fredenhagen-Roberts noncommutative spacetimes}
Canonical noncommutative spacetimes are characterized by a single
tensor $\theta_{\mu \nu}$, which of course takes different form
in different reference systems, breaking Lorentz symmetry.
It is possible to see the emergence of this $\theta_{\mu \nu}$
background as a result of a phenomenon of spontaneous symmetry breaking.
In fact, canonical noncommutative spacetimes could emerge from
more general spacetime theories in which the $\theta_{\mu \nu}$ tensor
is not fixed {\it a priori}.
A research line that originates in a 1994 paper by
Doplicher, Fredenhagen and Roberts~\cite{dfr}
is aiming for a theory in which $\theta_{\mu \nu}$ is itself associated
with a dynamical element of the theory.
Since no preferred $\theta_{\mu \nu}$ is introduced {\it a priori},
the theory is fundamentally Lorentz invariant (in the ordinary, undeformed,
sense). As usual, the fact that the dynamical equations of the theory
enjoy a certain symmetry, in this case Lorentz symmetry, does not
imply that the solutions of the theory be Lorentz invariant.
It is plausible that the ``vacuum" of the theory would be characterized
by a specific tensor $\theta_{\mu \nu}$ (but the Lorentz invariance of the
theory would then impose a large degeneracy of this vacuum, since acting
with a symmetry on a vacuum one must find another vacuum),
in which case one would talk of spontaneous symmetry breaking.
It is even  conceivable that all physical states (not only the vacua)
would be characterized by a nonvanishing value of $\theta_{\mu \nu}$,
so that in each physically viable realization of flat spacetime
there would be a preferred class of observers, again a mechanism
of spontaneous symmetry breaking (although of novel type).

In these cases the theory, {\it i.e.} the equations of dynamics, would be
Lorentz invariant but the spacetimes predicted by the theory
would break Lorentz invariance. The theory would be
fundamentally Lorentz invariant,
but the energy-momentum dispersion relations (possibly different for
different particles) would never be of the type $E^2 = c^2 p^2 + c^4 m^2$,
signaling that the symmetry is spontaneously broken.
At the fundamental level the theory could only predict a general formula
for the dispersion relation, involving the dynamical
variable $\theta_{\mu \nu}$, then in a specific spacetime (possibly
an eigenstate of $\theta_{\mu \nu}$) the dispersion relation
would take a specific form (which would ``break" Lorentz invariance
in the sense discussed above).

Of course, a theory that hosts $\theta_{\mu \nu}$ as a dynamical variable
might also not have spontaneous Lorentz-symmetry breaking,
if, for example, ``the vacuum" of the theory was characterized by
the condition $\theta_{\mu \nu} = 0$ or ``the vacuum"
was obtained as a large ``democratic" superposition of states characterized
by all possible values of the $\theta_{\mu \nu}$ matrix.

\section{On the fate of Lorentz symmetry in loop quantum gravity}
The two ideas for a nonclassical description of
spacetime that are being extensively
considered in the quantum-gravity literature\footnote{As mentioned,
it is still not clear whether the quantum-gravity problem necessarily
requires a spacetime picture that is fundamentally
nonclassical in the sense here advocated.
In particular, within the popular string-theory approach
to the quantum-gravity
problem the underlying spacetime picture is still not fundamentally
nonclassical; in fact, in string theory among the admitted spacetime
backgrounds it is still possible to choose spacetimes that are
completely classical. In that case physical processes still occur
in a classical (background) spacetime arena, and
spacetime is only ``quantized"
in the sense that some new particles (notably, the graviton)
are allowed to propagate (and mediate gravitational interactions)
in this fundamentally classical spacetime.}
are the ideas of noncommutativity and of discretization.
The observations reported in the previous Section confirm that
the fate of Lorentz symmetry is naturally nontrivial (in one or
another way) in noncommutative versions of Minkowski spacetime.
Also the idea of spacetime discretization naturally leads to
the expectation of a nontrivial fate for the continuous
classical Lorentz symmetry, but discretization can be introduced in
spacetime physics in many ways and it is difficult to make
very general considerations.
As mentioned, a simple-minded rigid discretization of Minkowski spacetime
would clearly not be consistent with the
continuous classical Lorentz symmetry~\cite{thooftdiscrete},
but it is not {\it a priori} obvious that the same would
happen in more sophisticated ways to introduce discreteness in
spacetime structure.
In this Section I consider
the best developed approach to spacetime discretization:
the one that emerged from research work
on ``loop quantum gravity"~\cite{carloliving,lqga,lqgb,lqgc}.

The ``loop quantum gravity" approach is perhaps the most
ambitious of all quantum-gravity approaches. While this approach,
not unlike all other quantum-gravity approaces,
is not immune from the presence of ``conceptual shortcuts"\footnote{Readers
familiar with the ``loop quantum gravity" approach
will realize that, among other things,
I am here concerned with the fact that the
diffeomorphism invariance originally sought by this approach
has never been truly realized, since the approach still basically
requires an {\it a priori} space/time split. Only invariance
under 3-dimensional space diffeomorphisms is genuinely maintained.
Also misterious (and suspicious) is the role
that the ``Immirzi parameter"~\cite{carloliving,lqga,lqgb,lqgc}
plays in the formalism. Moreover, the lack of a natural scheme for
the introduction of nongeometric degrees of freedom ({\it e.g.}
Standard-Model particles) is of course cause of serious concern.}
it is the only approach that does not rely on an {\it a priori}
spacetime background.
Spacetime-background independence
is a very natural (but technically challenging)
requirement for theories, quantum-gravity theories,
attempting to address the ``conceptual
tension" between quantum mechanics and classical general relativity.

In fact, general relativity is a
background-independent description of spacetime dynamics.

Of course, for a background-independent approach an important
task is the one of describing
those physical contexts in which a background spacetime does
emerge\footnote{The majority and the simplest
of our observations are naturally described
as processes occurring in a classical (and often nearly flat)
spacetime arena, but the loop quantum gravity approach is still
unable to describe that simplest type of phenomena.}.
This task has not yet been accomplished in loop quantum gravity,
but progress in this direction might be forthcoming.
In general it is not surprising that the analysis of this ambitious theory
turns out to be extremely difficult. Only very few characteristic
predictions have been obtained. Among these predictions the most
celebrated are the ones concerning some natural area
and volume ``observables" (here intended in the technical/mathematical
sense) and the spectrum of these observables, which turns out to be
discrete. The discretization of the spectrum of geometry/spacetime
observables is the most fundamental characterization of
a discretized picture of spacetime.

I shall not review here the loop quantum gravity approach, not even
for what concerns the derivation and analysis of the area and volume
operators. These topics are very effectively and pedagogically
reviewed in some recent publications (see, {\it e.g.},
Refs.~\cite{carloliving,lqga,lqgb,lqgc}).
For the purposes of my analysis it is sufficient to comment
here on some qualitative aspects of those results.
I am in fact exclusively interested in the way in which
discretization of areas and volumes might affect Lorentz symmetry.

In spite of the fact that the classical-Minkowski limit has not yet
been found in loop quantum gravity (actually at present the programme
is still attempting to identify a suitable limiting procedure
for the emergence of classical spacetimes;
see, {\it e.g.}, Ref.~\cite{fotiniNEWclassicallimit})
this concern about Lorentz symmetry is not premature
since the area/volume discretization results are understood
as completely general: those discretizations are a general
prediction of the loop quantum gravity approach, which should in
particular apply to the zero-curvature (Minkowski/quasi-Minkowski) limit.
While a dedicated study of this issue is still lacking,
at conferences and from the introductory remarks of review papers
it appears that two intuitions are emerging:
according to one of these intuitions one expects that, in spite
of the worrysome appearance, these discretizations should turn out
to be compatible with classical ordinary Lorentz symmetry,
because at the level of the tools introduced in the formalism
it appears that nothing could have spoiled the symmetry,
while the other intuition assumes that it would be impossible
to reconcile the discretization of these spectra
with continuous (classical) transformations of Lorentz type.

I argue here that both intuitions rely on false premises.
The intuition which is favourable to the survival
of classical ordinary Lorentz symmetry focuses exclusively on properties
of the formal tools introduced in the formalism
(while, as emphasized in the previous Sections,
it is at the level of the physical predictions
that Lorentz symmetry should be analyzed)
and neglects some key differences
(see below) between space rotations
and Lorentz boosts that are relevant for a canonical
quantum theory.
The intuition which is hostile to the survival
of classical ordinary Lorentz symmetry assumes that discrete spectra
are inevitably inconsistent with the presence of continuous
classical symmetries, while the case of classical space-rotation
symmetry in ordinary quantum mechanics (here reviewed in Section~2)
shows that there are counter-examples for this,
however intuitive, argument.

Some insight on this delicate (and crucial, especially considering
the mentioned developing experimental situation)
issue can be gained using as guidance the comparison between
the description of angular-momentum discretization
in ordinary quantum mechanics
and the description of area/volume discretization in loop
quantum gravity.
As clarified in Section~2,
angular-momentum discretization
in ordinary quantum mechanics is consistent with
classical space-rotation symmetry because the $L^2$ discretization
involves an observable on which the classical symmetry acts trivially
(an invariant), and $L_x$ discretization (just like the discretization
of $L_y$ and $L_z$) involves an observable on which
the classical symmetry does not act at all.\footnote{As emphasized
in Section~2,
when the observer $O$ measures $L_x$ in a space-rotation invariant
theory it is still not possible to make a definite prediction
for the components $L_{x'}$, $L_{y'}$, $L_{z'}$ that are relevant for
another observer $O'$. In order to predict $L_{x'}$
(and/or $L_{y'}$ and/or $L_{z'}$) one must do three measurements:
$L_x$, $L_y$ and $L_z$.}
As emphasized in Sections~3 and 4, Lorentz-symmetry transformations
do not act on the area\footnote{Since my remarks apply equally both to
the area and the volume operators, for short I will often refer
only to the area operator.}
observable $A$.
If the observer $O$ measures the area $A$ of the surface of a table,
but does not measure the velocity $V$ of the table, Lorentz symmetry
is unable to predict the size of that area according to another
observer $O'$ moving at velocity $V_0$ with respect to $O$.
If instead the observer $O$ measures {\underline{both}}
the area of the surface of the table and its
velocity, a $(V,A)$ measurement,
then Lorentz symmetry makes a definite prediction:
it establishes that (assuming velocities are all
alligned) $(V,A)=(V,\sqrt{1-V^2/c^2} A_0)$, where $A_0$
is the rest area of the surface of the table
and Lorentz symmetry also establishes that,
denoting with $V_0$ the relative $OO'$ velocity,
$A'=\sqrt{1-V'^2/c^2}A_0=\sqrt{(c^2-V'^2)/(c^2-V^2)}A$,
where $V'(V_0,V)$ is the velocity of the table with respect to $O'$.
So Lorentz symmetry is fully operative on the $(V,A)$ measurement,
while it bears no relevance when only $A$ is measured.

Lorentz-symmetry transformations
act on the double measurement $(V,A)$ in the same sense that
space-rotation-symmetry transformations act on
the triple measurement $(L_x,L_y,L_z)$.
Just like classical space-rotation symmetry is consistent with
$L_x$ discretization if and only if (see Section~2)
the theory also predicts
that a sharp measurement of $L_x$ imposes a large uncertainty
on $L_y$ and $L_z$,
ordinary classical Lorentz symmetry is consistent with
$A$ discretization if and only if the theory also predicts
that a sharp measurement of $A$ imposes a large uncertainty
on $V$.

This point I am raising concerning the $(V,A)$ measurement
and the fact that area discretization is not necessarily in
conflict with Lorentz symmetry should be distinguished from
a previous argument, due to Snyder~\cite{snyder},
also relevant for the possible compatibility
of Lorentz symmetry with spacetime discreteness.
Snyder proposed and analyzed a specific type (with coordinate
commutators not expressable as functions of the coordinates themselves)
of flat noncommutative spacetime, finding that Lorentz
symmetry was maintained and that the commutation
relations would induce coordinate discreteness (which, I observe,
should be discussed cautiously, in terms of distances
and other diffeomorphism-invariant entities).
In the Snyder spacetime
the sharp measurement of one of the coordinates would in general
impose an uncertainty on the time coordinate (and the other coordinates).
One clear difference is that my argument concerns area
(and associated surface velocity) measurements,
while Snyder's concerns space coordinate (and associated time coordinate)
measurements. A more subtle, but perhaps even more important,
difference is that Snyder only verified Lorentz symmetry as a
property of the formalism: Snyder introduced Lorentz-symmetry
transformations as
a set of transformations for the ``coordinate observables",
and only emphasized that the spectra of the coordinate
operators are the same for all observers.
Instead here I discussed Lorentz symmetry at the level of
the predictions of the theory, and in particular I emphasized
that some of the measurement procedures that are governed by
Lorentz symmetry can be shared by different observers
(see Subsections 3.3, 3.4 and 3.5) and that this might render
insufficient any general argument on the spectra of symmetry-connected
observables (one might have to perform a dedicated analysis of
the action of Lorentz symmetry on eigenstates of some of the
observables).

From these observations it follows that area discretization does not
in itself represent a violation
of ordinary classical Lorentz-symmetry, and therefore
in each quantum-gravity approach predicting area discretization
a dedicated analysis of the fate of Lorentz symmetry is needed.
I shall argue that the present level of development of loop
quantum gravity is insufficient for obtaining a definite
answer to this important question, especially in light of
the fact that an observable ``area of the surface" was introduced
without the introduction of an observable ``velocity of the surface",
which, from a Lorentz-symmetry perspective, should naturally accompany
the surface area observable.
From my analysis it emerges that a natural path for
the realization of ordinary Lorentz symmetry in loop quantum gravity
can be found through the analogy of the status of space-rotation
symmetry in quantum mechanics. In order to achieve this goal
a suitable observable ``velocity of the surface" must be found,
and some of its relevant properties must be analyzed.
In particular, one should find that
in surface-area eigenstates the velocity of the surface
is affected by a large uncertainty (large enough to comply with the
requirements sketched in Subsection~2.4).
But even if this task was accomplished ({\it i.e.} if indeed the
analogy between space-rotations in quantum mechanics
and Lorentz transformations in loop quantum gravity was successful
in the sense of finding satisfactory commutation relations
between surface velocity and surface area)
the status of Lorentz symmetry in loop quantum gravity would
still require further investigation.
In fact, from the perspective of a canonical quantum theory,
such as the loop quantum gravity here considered,
the analogy between space-rotation symmetry and Lorentz symmetry
has several limitations.
The special role that time plays in a canonical theory
imposes that one should consider observers connected by a
a Lorentz boost in a way that is significantly different
from the one appropriate for observers connected by a
space rotation: one can roughly say that a space rotation
acts within a given Hilbert space (a given choice of the
time variable) whereas Lorentz boosts connect a given canonical
theory (a given Hilbert space and time variable)
with another {\underline{different}} canonical theory.
In light of this difference it is not even
conceivable that the status of Lorentz symmetry in
a canonical theory could be truly analogous to the
status of space-rotation symmetry in a canonical theory.
On the contrary it appears plausible that at some level
of analysis the fact that the canonical theory is forced
to make direct/primitive reference to equal-time surfaces,
rather than to the surface's world sheet,
might have significant implications.
As emphasized in the previous sections,
in the preservation of space-rotation symmetry in quantum mechanics
a key role is played by the fact that the theory and our description
of measurement procedures are most primitively formulated
in terms of the physical entities, like the angular-momentum vector,
whose objectivity is codified by space-rotation symmetry transformations.
The observable simply denoted by ``$L_x$" in the formalism
inevitably corresponds physically to an observable obtained from
two ``objective/physical vectors", the angular-momentum vector $\vec{L}$
and a second ``projecting" vector $\vec{B}$.
Meaningful measurement procedures
can be devised when the value of $\vec{B}$ is known.
A faithful description of Lorentz symmetry might require that
the theory be formulated in terms of the world-sheet (world-line,
world-volume,...) and that the observable equal-time area
be introduced as an appropriate ``projection" of the world sheet
(a ``projection" that identifies the time variable with respect
to which one is defining equal-time surfaces).
A conventional canonical quantum theory cannot be formulated in
this way, since it requires an {\it a priori} time variable/parameter.
This problem of the ``absence of the world-sheet"
might already be hidden in the nature of the ``area observable"
presently adopted in loop quantum gravity; in fact that observable
refers directly to an equal-time coordinate area (although it is
clear to loop-quantum-gravity experts that the surface
should be eventually meaningfully identified through some fields/particles)
and it is not seen as some projection of a world-sheet: it is indeed
like introducing in ordinary quantum mechanics
as most primitive observable $L_x$, without any room for
identifying this $L_x$ as a projection of the type $\vec{L} {\cdot} \vec{B}$.

I shall also argue that the problem of the ``absence of the world-sheet"
is also significant from the perspective of transforming the formal
result of a discrete spectrum for the area observable into a
physical prediction. A necessary (actually not sufficient)
condition for theorists to make a physical prediction of discretization
is to indicate at least one measurement procedure in which
this discretization is (at least in principle) observable, and analyze
the chosen measurement procedure within the adopted theory in order
to verify that the full theory is consistent with the observability
of the discreteness. Since loop quantum gravity is not formulated
in terms of world sheets one runs into a situation which
is alarmingly different from the one in which one endows
angular-momentum discretization with the status of a genuine
physical prediction: for example, the Stern-Gerlach device realizes
physically the projection (also allowed/admitted by the formalism)
of the vector $\vec{L}$ along the direction
of another vector (the direction of the magnetic field $\vec{B}$
which characterizes the measurement procedure).
So it appears that no specific measurement procedure can be suggested
by the formal result, and, in addition, I shall show
that the analysis of some standard/familiar
area-measurement procedures suggests a paradoxical situation
in which area discretization would
not be observable, even if rigorously introduced at the level of
formalism.

My observation that, from the perspective of a canonical quantum theory,
such as the loop quantum gravity here considered,
the analogy between space-rotation symmetry and Lorentz symmetry
can only be adopted in a partial/limited way is also based
on some aspects of the measurement procedures we
presently adopt to give operative meaning
to the relevant observables.
I must stress here again
that some of the measurement procedures that are governed by
Lorentz symmetry can be shared by different observers
(see Subsections 3.3, 3.4 and 3.5) and that this might render
insufficient any general argument on the spectra of symmetry-connected
observables.
In the length measurement of Subsection~3.4 both observer $O$
and observer $O'$ measure time (which is there used to measure length)
using {\underline{the same}} clock.
In the analysis of that length-defining procedure (and of analogous
procedures for the measurement of other spacetime entities,
like areas) it appears impossible to contemplate the possibility
that observer $O$ would characterize the situation with a length
eigenstate while the observer $O'$ characterizes the situation
with a superposition of length eigenstates: both observers
establish whether or not they are dealing with a length eigenstate
using {\underline{the same}} clock readouts!
This appears to be different from the case of measurement procedures
for angular-momentum components: a Stern-Gerlach device measures
one and only one component of angular momentum. (Instead the length-measuring
setup of Subsection~3.4 measures both the ``equal-time
area projection to observer $O$ of the surface's world-sheet"
and the ``equal-time
area projection to observer $O'$ of the surface's world-sheet".)
I shal also emphasize another difference: while the value of the, say,
$L_y$ component of angular momentum is not relevant for
procedures that measure $L_y$, and therefore the noncommutativity
of $L_x$ and $L_y$ does not affect the analysis of the measurement
procedure (it only affects the logical structure of the results of
different measurement procedure),
the knowledge of the velocity of the surface appears to be needed
in order to introduce meaningful area-measurement procedures,
and therefore a noncommutativity of surface-velocity
and surface-area would have severe implications
for the analysis of area-measurement procedures.

These remarks set the agenda for my analysis of the status
of Lorentz symmetry in loop quantum gravity, an admittedly rich
agenda.
The delicate nature of some of the points I intend to raise
forces me to organize the remaining subsections of this section
in a way that does not have a nice flow. Rather than recognizing a
logical order in the sequence of the subsections the reader should
attempt to recognize that most subsections provide
material in support for the remarks anticipated in these
opening paragraphs of this section devoted to loop quantum gravity.
In addition, some of the subsections report observations that are
not directly relevant for the line of reasoning I proposed in
these opening remarks, but instead they intend to provide to the
reader some material useful for comparison with other recent studies.
I start with some general remarks on the way in which a meaningful
investigation of Lorentz symmetry can be performed in a dynamical
(and quantum) theory of spacetime.

\subsection{Investigating Lorentz symmetry in dynamical quantum
theories of spacetime}
The focus of this paper is global Lorentz symmetry, which can be
rigorously investigated in the nondynamical flat
noncommutative spacetimes considered in the preceding section,
but is not a natural element of a dynamical quantum theory
of spacetime. Global Lorentz symmetry does not even have a
a truly fundamental
role in classical general relativity, where it only emerges
on what should be seen as very special solutions of the dynamical
equations (or as an approximate symmetry of spacetimes
that are well approximated by Minkowski in a region
of small size). The role of Lorentz symmetry is likely to be
even less on the forefront of the structure
of a quantum dynamical theory of spacetime.
However, a quantum-gravity theory should be able to describe
the contexts with which we are observationally familiar, in which
spacetime is to a good approximation flat and classical.
Most of my considerations concern
these quasi-Minkowski spacetimes, which must be admitted
by quantum-gravity theories. Although they might play a relatively
marginal role in the conceptual/technical structure of quantum-gravity
theories, these spacetimes are likely to provide our best
experimental-testing ground for
quantum-gravity theories~\cite{polonpap}.

Even in a ``quasi-Minkowski" spacetime one might wonder whether
Lorentz-symmetry transformations (or some predictable departure
from their structure) should play a role in quantum theories of
spacetime. However, I am adopting here the working assumption
that a meaningful test of Lorentz symmetry should be possible
in quantum gravity. A quantum-gravity theory must have a
classical-spacetime $L_p \rightarrow 0$
limit and a further zero-curvature $R \rightarrow 0$
limit should introduce a corresponding
role for Minkowski spacetime in quantum gravity.
Of course, in that limit, which I am here schematically describing
as a $L_p \rightarrow 0$, $R \rightarrow 0$ limit,
ordinary Lorentz symmetry should hold.
If then the theory allows us to consider (both formally and
operatively/experimentally) situations in which the
$R \rightarrow 0$ condition is maintained but
the $L_p \rightarrow 0$ condition is softned ($L_p$ small but nonzero)
there are basically two possibilities: either Lorentz symmetry
still holds exactly or there are small ($L_p$-suppressed)
departures from Lorentz symmetry.
The analysis I am reporting in this paper hopes to provide useful
elements for the investigation of quantum theories of spacetime
in this respect, for establishing whether $L_p$-suppressed departures
from Lorentz symmetry are to be expected.

\subsection{Procedures for area measurement}
Since I am here considering the status of
Lorentz symmetry in loop-quantum-gravity,
the first point I must be concerned with is
whether the formal result of area discretization
can be adopted as a truly physical prediction.
In loop quantum gravity there is indeed an operator which in the
classical limit represents the area of a surface and at the quantum
level turns out to have discrete eigenvalues.
Does this quantum operator represent physical areas also
at the quantum level?

It is of course not sufficient to call an operator ``area" for it
to qualify as the description of physical areas.
We should at least specify one class of area-measurement procedures
for which we predict that the outcome of the measurements
would reflect the discretization.
We should also analyze the proposed measurement procedure
and verify that there are no logical obstructions for that
discretization to be revealed. (I stress that this is not a point
about our {\underline{technological ability}} to reveal the discretization:
in order to formulate a physical prediction it is sufficient
that the adopted measurement procedure does not have
any {\underline{in principle}}
obstructions for revealing the discretization.)

The prediction of $L_x$ discretization in ordinary quantum mechanics
is more than a formal result: it refers to specific angular-momentum
measurement procedures, of which the Stern-Gerlach procedure is a prototype.
The interpretation of angular-momentum discretization
within ordinary quantum mechanics
is very simple: the same observable quantity
({\it i.e.} the quantity measurement through the same operative procedure)
that is called angular momentum in classical mechanics also exists
in quantum mechanics and it can be measured exactly in the same way,
but according to quantum mechanics the results of these measurement
procedures, which in classical mechanics could take any arbitrary value,
can only take certain discrete values.
Also in classical physics we could use a Stern-Gerlach-type device
to measure the angular momentum of a
charged spinning ``classical particle" (small ball).
The loop-quantum-gravity
results on area/volume discretization
are presently being discussed as if they were to be interpreted in
complete analogy with the known angular-momentum discretization,
as if we should be able to adopt the same operative definition
of area we presently adopt, the same area-measurement procedures
we presently adopt, and for those measurement procedures the theory
would predict discrete outcomes.

Angular-momentum discretization is a general prediction of ordinary
quantum mechanics. It applies also to macroscopic systems, but there
we lack the needed experimental accuracy to reveal it.
If the loop-quantum-gravity area discretization is to be interpreted
just as the angular-momentum discretization in ordinary quantum
mechanics it should mean that area-measurement procedures
performed on macroscopic surfaces (macroscopic with
respect to the Planck distance scales) should also give
sharp discretized outcomes, which we would have not noticed because of
lack of the needed experimental accuracy.

But I conjecture that the area operator of loop quantum gravity cannot
be interpreted as a new description of the same area observable
we are all familiar with in classical physics, and that the
loop-quantum-gravity discretization of area is {\underline{unobservable}}.
I repeat: here I do not mean that, because of the small discretization
scale, we might never be able to reveal the effect. I mean that the
the theory itself should predict that this discretization
cannot be observed, that there be an in-principle obstruction to
its observation.
If this conjecture is correct, the associated reanalysis of the area
and volume operators might also lead to a reassessment of the
status of Lorentz symmetry.

My conjecture is based in part on the fact that the same physical
intuition which motivated the description of areas at the formalism
level in terms of a quantum operator should also motivate a reanalysis
of area-measurement procedures, and this reanalysis suggests
that there should be an absolute limitation on the measurability
of areas. This sort of new uncertainty principle would also
render unobservable, and therefore unphysical, area discretizations
of the type discussed in loop quantum gravity, with Planckian
area quantum.

In building up to this intuition let me start with a
given procedure for the measurement of areas.
Unfortunately, in spite of the relevance of the area results
at the formal level, and therefore the need to endow with physical
meaningfulness those results through a defining measurement procedure,
not much has been done on this point in the loop-quantum-gravity
literature. One noticeable exception is the deservedly popular
study reported in Ref.~\cite{carloarea}.
The measurement procedure there adopted is relevant for the case in which
the matter fields that specify the surface whose area is being measured
are taken to form a metal plate, and the area $A$ of this metal plate is
measured using an electromagnetic device that keeps a second metal
plate at a small distance $d$ and measures the capacity $C$
of the capacitor formed by the two plates.
The primary measurement would be the capacity, and the sought
area would be evaluated through the relation\footnote{In
Eq.~(\ref{adc}) the presence of $\epsilon_0$ reflects the
simplifying assumption that the measurement be perfomed in
absolute vacuum. This simplification does not affect
the validity of my remarks.}
\begin{eqnarray}
A = {d \over \epsilon_0} C ~.
\label{adc}
\end{eqnarray}

This is as good as any other candidate area-measurement procedure
on which to explore the physical interpretation
of the loop-quantum-gravity area-discretization result.
It could be for area discretization in loop quantum gravity
the analog of what the
Stern-Gerlach setup is for angular-momentum discretization
in ordinary quantum mechanics.
In the Stern-Gerlach setup one measures the angular momentum
through the measurement of the position of arrival
of the particles on a screen/detector, and the prediction
is that those positions take discretized values, reflecting
angular-momentum discretization.
In the area-measurement procedure considered in Ref.~\cite{carloarea}
one measures the area through the measurement of the
capacity $C$.
Is then the prediction of the loop-quantum-gravity approach
that those capacity readouts can only take certain
discrete values?
This is one of the hypotheses raised in Ref.~\cite{carloarea},
but in this respect several considerations are in order.

First of all let me emphasize that of course
this area-measurement setup cannot follow instantaneously
the time evolution of the surface (and of its area).
Typically the time of measurement required by area-measurement
is at least of order $T \sim \sqrt{A}/c$. For this area-measurement
procedure
based on capacity measurement this estimate of the time of measurement
is found by considering the time needed by the capacity to respond to a
sudden change in the area of the plate being measured.
After such a sudden transition from one size of the area to another
the distribution of electrons on the surface of the
metal plate will have to readjust over the whole surface until
a new equilibrium is reached, and this transition from the
previous equilibrium configuration to the new one must take
at least a time of order $\sqrt{A}/c$ (since $c$ is the limiting
speed for information).
In a theory where area eigenstates are states in which the metric
is sharply defined, if the theory also predicts that correspondingly
the time derivate of the metric is largely uncertain,
this area measurement setup would not be able to give
the value of the area of the surface at a certain specific
time $t^*$ but would rather give the average size of that
area over a time interval of order $\sqrt{A}/c$.
The net result would be that there would be no trace of the
discretization.

Also notice that the formula (\ref{adc}) which is at the core
of this measurement procedure (just like the relation between
external magnetic field, relevant component of the angular momentum
and point of arrival on the screen is at the core of the Stern-Gerlach
procedure) implicitly assumes that the surface be absolutely flat.
In a theory in which area eigenstates were not allowed to be exactly
flat the discretization of the area would be masked (hidden) by the
fact that this measurement procedure makes some averaging
over the fluctuations with respect to exact flatness that
the quantum theory predicts.

Also notice that this measurement procedure basically assume
that one can measure accurately the velocity of
the (metallic) surface whose
area is being measured. In fact it appears to be necessary to assure
that the two surfaces that compose the capacitor are parallel
(constant distance $d$) and that they be centered with respect to one
another\footnote{If the second surface (the one that belongs to the measuring
device) is much larger than the surface whose area is being measured
one should be concerned about ``boundary effects" since the
formula (\ref{adc}) actually assume a highly symmetric configuration
(it strictly applies to infinite parallel metallic plates).
If the two surfaces are roughly of the same size any relative velocity
would of course affect the capacity. Moreover, for a charged metallic
plate which is not at rest one should worry about associated
magnetic fields.}.

Finally let me make a point which I already
stressed in Refs.\cite{areaarea}.
In this area-measurement procedure
based on capacity measurement it is necessary to measure the distance
between the plates.
If $d$ is not known sharply then the relation between $C$ and $A$
becomes fuzzy and the discretization of $A$ may become unobservable.
This observation appears to be relevant for theories, such
as loop quantum gravity, in which there appears to be a well-defined
area operator but some difficulties are encountered in describing
distances.
It would be paradoxical~\cite{areaarea}
for a theory predicting that distances
cannot be measured with absolute accuracy
to predict that the results of this capacity-based area-measurement
procedure should be sharp (and discretized).

\subsection{Toward a loop-quantum-gravity
description of fuzzyness}
As I hope it emerged from the line of analysis advocated
in this paper,
if it was true that the measurement of a flat area described
in the preceding subsection would give discretized results
and one was able to measure rather accurately the velocity
of the surface (as it appears to be required by that measurement
procedure) then Lorentz symmetry should be necessarily violated
as explained in the preceding sections.
My conjecture is that more refined analyses of the
loop-quantum-gravity formalism should find that
the discretization of areas is not a physical prediction of
the formalism, that it would not show up in any
procedure for the measurement of areas.
If such developments did come about one should then
reassess the status of Lorentz symmetry in loop
quantum gravity.
I am raising the possibility that
the present loop-quantum-gravity
results on area/volume discretization should not be intended as a genuine
discretization (at least not in the sense we presently understand
angular-momentum discretization).
They should rather be an awkward way in which the theory
renders us aware of a new absolute limitation
on the accuracy with which areas can be measured.

The other possibility (which I find to be less likely, but is
certainly plausible)
that emerges from this study is that instead
area discretization is indeed a physical prediction of
loop quantum gravity and Lorentz symmetry is preserved, but,
in light of the arguments here presented, in this case it is
inevitable that the formalism should also predict that the
surface velocity is largely undetermined on surface-area eigenstates
and in addition one should face the challenges of:
(i) identifying a area-measurement procedure which is not affected
by the surface-velocity uncertainty, and (ii) understand the
role that measurement procedures such as the ones described
in Subsections~3.3, 3.4, 3.5, in which the readouts
of the same measurement procedure are meaningful for different
inertial observers, should have in the theory (they should
somehow become disallowed, otherwise they would impose that
area eigenstates are area eigenstates for all observers,
with the consequence, in which case the discretization
scale of the area spectrum could not be observer-independent.

I will, for short, refer to these two possibilities
as the ``fuzzyness" scenario and the ``area/velocity
noncommutativity" scenario.

In the previous subsection I have provided some physical arguments,
through the analysis of a measurement procedure, that support
my conjecture
in favour of the ``fuzzyness" scenario.
Here I want to provide some remarks on the formalism
that go in the same direction.

A first point that needs careful consideration
is the fact that the area ``measured" by the
loop-quantum-gravity area operator is
an area defined by corresponding conditions on coordinates, rather
than the area of a surface
physically/meaningfully identified by some field/matter distributions.
Some work in the direction of such a meaningful identification
has been done, but the problem appears to be dangerously entangled
with the general open problem of loop quantum gravity for
what concerns the introduction of nongravitational degrees
of freedom (such as realistic descriptions of the standard
model of particle physics).
Lacking this technical ingredient one might be tempted
to adopt the viewpoint in which a surface is
meaningfully identified by some conditions
on the boundary coordinates, if these coordinates are intended
as physical distances from the axes of the ``laboratory" of an
observer. However, this
interpretation is, in my opinion, rather troublesome when the tetrads
are promoted, as done in loop quantum gravity, to the status of quantum
variables. This should intuitively mean that, in a given ``state of
spacetime" the laboratory
axes (the laboratory frame) cannot be identified with the same
sharpness as in classical physics. One might end up evaluating
the sharp spectrum of a ``coordinate area" in a framework where
the physical meaning of those coordinates is not itself sharp.
Even the ``coordinate area", with coordinates intended
as physical distances from the natural reference axes of a laboratory
of course should involve some suitable field/matter distributions
that identify those axes. As observed in Refs.~\cite{pap16},
reference axes can be physically identified with absolute precision
in classical physics, where one could for example use particles of
negligibly small mass rendering gravitational evolution negligible.
Reference axes can also be physically identified with absolute
precision in ordinary (no-gravitation) quantum mechanics, where one
would instead choose to identify the axes using very massive particles
for which the uncertainty principle is negligible. But in quantum
gravity axes identified by very massive particles would have ``problems"
due to gravity and axes identified by nearly-massless particles
would have ``problems" due to the uncertainty principle.
This line of reasoning provides support for the ``fuzzyness"
scenario even when not focusing on the nature of
material reference systems: in fact the argument
also applies to the analysis
of a ring of particles used to identify a surface.
In order to measure the area of that surface we need that
the particle be in rigid motion for the time needed to complete
the measurement procedure, which should be at least
a time of order $\sqrt{A}/c$. In order to suppress the gravitational
interactions among the particles (which could lead to area deformation
on time scales shorter that the time of measurement)
one would like particles of very small mass, but then Heisenberg's
uncertainty principle would introduce a large uncertainty in the motion
of the particles and the area would vary on very short time scales anyway.
If we take particles of large mass the uncertainty principle would
not cause problems, but rigid motion would be in conflict with
the strong gravitational fields being generated by the massive
particles. So in a quantum theory of gravity a surface which is
physically/meaningfully identified by a ring of particles
will be deformed (at a level significant for Planck-scale accuracy
of area measurement) on time scales that are shorter than
the time needed by the measurement.

Another issue that should be addressed in the formalism is
the one concerning the role that the time derivatives
of the metric play in the procedures used to give operative
meaning to the concept of area.
One could study how the analysis of Lorentz symmetry
is affected by the fact that the concept of area of a surface
involves the metric,
while the concept of velocity of an area involves (in an
appropriately weak sense, see below)
the time derivatives of the metric.
In a theory predicting nontrivial commutation relations between
the metric and its derivatives this might generate
the type of surface-velocity/surface-area noncommutativity
which I have shown here to be needed in order to have
a Lorentz-symmetric description of area discretization.
This possibility certainly deserves detailed investigation.
It is probably reasonable to conjecture that
nontrivial commutation relations between
the metric tensor and its time derivative will introduce some
level of noncommutativity between the velocity of the area
and the area itself, but probably the induced velocity
uncertainty in area eigenstates would not be large enough to rescue
classical continuous Lorentz symmetry (not large enough
in light of the argument here discussed
in Subsection~2.4).
The point is that the time derivatives of the metric tensor
do not properly contribute to the velocity of the area in the sense
relevant for Lorentz-symmetry transformations.
The relevant physical property described by Lorentz symmetry
concerns measurements of rigidly-moving surfaces (surfaces whose
area does not change significantly during the time of measurement)
in which one simultaneously measures the area and the velocity
of the surface. In Minkowski spacetime it doesn't matter which point
on the surface we choose in defining the velocity of the surface.
At the instant $t^*$ at which the measurement is performed
one is free to choose this point for the identification of the
velocity as the origin of the reference
system adopted by the inertial observer, and in this case
the time derivatives of the metric tensor
should not contribute significantly to the velocity of the area.
They rather contribute to the
deformation of the surface as a function of time
(a sort of velocity of deformation of the surface,
velocity that describes how quickly the area changes in time).
For what concerns Lorentz symmetry,
the most important consequence of
nontrivial commutation relations between
the metric tensor and its time derivative should be a possible
absolute limit on rigid motion of a surface.
This also fits the intuition~\cite{pap16}, described above,
that in a theory in which both
the uncertainty principle and gravitational interactions are
included, as expected for quantum-gravity theories, rigid motion
would not be allowed.
The study of the time derivatives of the metric tensor
might simply allow us to establish that, as conjectured above,
the area spectrum derived
in loop quantum gravity would not describe a new
phenomenon of discretization of the familiar concept of area,
but would rather reflect an absolute limitation on our
present idealization of the operative definition of the area of a surface
of a moving body.
Since it appears inevitable that the measurement of an area of
(roughly determined)
size $A$ will require at least a time
of measurement $T_{meas} = \sqrt{A}/c$,
if areas cannot move rigidly ({\it i.e.} if the theory predicts
that the area of a surface must vary on very short, perhaps Planckian,
time scales) how well could we measure them?
If the spectrum is quantized with $L_p^2$ discretization scale
would we be able (even at the {\it gedanken-experiment} level)
to measure the area accurately enough to find evidence of this
discretization? If not,
in which sense would the discretization be ``real"?

Again on the point of the noncommutativity between metric and
its derivatives I want to observe that this should most intuitively
contribute to some sort of uncertainty between the shape of the
boundary of the surface (which specifies the area)
and the time variation of that shape (which specifies a limit
on rigid motion). Instead the compatibility of Lorentz symmetry
with area discretization requires,as here shown, a certain
type fo noncommutativity between surface velocity and surface
area, which raises a few puzzles. One of these puzzles if that
we already know that the velocity of the surface will become
totally uncertain if we measure the position of the table
(this follows strightforwardly from ordinary quantum mechanics),
and now it might be surprising (and perhaps hard to implement
technically) if the uncertainty of the velocity of the surface
is also subject to limitations linked to the uncertainty
in the area of the surface.\footnote{Here I am puzzled by the
fact that usually we see the position and velocity of the
surface as properties of the COM system, while area and positions
of the boundary points are seen as properties of the motion
relative to the COM. The two aspects of the problem usually decouple
and instead the consistency of Lorentz symmetry with
area discretization appears to require that there be a link
between the uncertainty of the area (a property of the
motion of the boundary points with respect to the COM)
and the uncertainty of the velocity of the surface
(a property of the COM).}

Concerning procedures for area measurement,
as mentioned, besides the issue of the needed time of measurement
it might also be significant the role that length measurements
have in the measurement of areas,
and this in turns provides me another link between properties
of the loop-quantum-gravity formalism and the ``fuzzyness" scenario.
While the discrete spectrum of the area and volume operators
is (in the sense here discussed)
a well-established prediction of the loop-quantum-gravity
formalism,
the status of the length observable in loop quantum gravity
is still unsettled~\cite{carloliving,lqga,lqgb,lqgc}.
It is difficult~\cite{areaarea}
(perhaps impossible) to devise an area-measurement
procedure that truly avoids the use of
length/distance measurements.
If in loop quantum gravity the concept of length
turned out to emerge as ``inherently fuzzy"
this would then affect
any area measurement that involved a length measurement:
this type of area measurement would be
subject to the fuzzyness of lengths, and if the fuzzyness scale was larger
than the area discretization scale the discretization would
become unphysical/unobservable~\cite{areaarea}.

Looking at
the type of spacetime geometries that correspond to
area-operator eigenstates one can find additional encouragement
to interpret area/volume discretization
in loop quantum gravity as a manifestation of a new limitation
on the measurability of area,
rather than as a genuine discretization of the results of the
area-measurement procedures we presently use to define
areas operatively.
As discussed, {\it e.g.}, in Ref.~\cite{carloliving},
these spacetime geometries do not look anything like a quasi-classical
spacetime. At present the concept of area is well understood in classical
spacetime. Quantum-gravity theories should first of all tell us how
the concept of area can be introduced (and the properties it acquires) in
spacetimes which are nearly, but not exactly, classical.
Area-operator eigenstates do no admit this interpretation.
A compelling quasi-classical (and particularly quasi-Minkowski)
limit has not yet emerged in the quantum-gravity literature, but it
appears likely that these quasi-classical spacetimes would be described
in terms of a large superposition of area eigenstates. Measuring area
in such quasi-classical geometries, the only ones that are likely to
be accessible to us, we would never find evidence of area discretization,
but only of some sort of fuzzyness of the area. Again this argument
suggests that the discretization that emerged in the loop-quantum-gravity
literature might not have
the same physical meaning as other, more familiar, examples of
discretization, such as angular-momentum discretization.
When we talk of angular-momentum discretization we are still describing
angular momentum in the same way as done in classical phyiscs, and we
are introducing a new property (discretization) without changing the
classical concept of angular momentum.
The loop-quantum-gravity area discretization might instead not
admit interpretation as a prediction
of discrete results for measurements
of the familiar area observable, but rather a manifestation of the
fact that at the quantum-gravity level that simple-minded concept
of area (as presently defined at the operative level)
is no longer applicable.

\subsection{On the role of the surface velocity in
area-measurement procedures}
Some of the points I raised on the loop-quantum-gravity area operator
came from the observation that Lorentz symmetry acts on the $(V,A)$
measurement (simultaneous measurement of the area and the velocity of
a rigidly-moving surface), just like space-rotation symmetry acts
on the $(L_x,L_y,L_z)$ measurement (simultaneous measurement of the three
components of angular momentum). In this Subsection I want to raise the
possibility that connection between $V$ and $A$ measurements in a
$(V,A)$ measurement might be stronger than the connection between
$L_x$, $L_y$, and $L_z$ measurements in a
$(L_x,L_y,L_z)$ measurement.

For the point raised in this Subsection a key ingredient of
the Stern-Gerlach procedure is that the corresponding $L_x$ measurement
does not depend on the values of $L_y$, and $L_z$:
in the Stern-Gerlach procedure by measuring
the point of arrival of an electron on a screen one can
reliably deduce
the value of $L_x$ independently of the values of $L_y$ and $L_z$.
The fact that $L_y$ and $L_z$ are not known does not affect the
precision of the $L_x$ measurement which in fact can be absolutely
sharp (in principle).
The point is that in the Stern-Gerlach setup the equation that
translates the measured ``point of arrival of the particle on the screen"
into an $L_x$ measurement does not depend in any way on the values
of $L_y$ and $L_z$, so the measurement of $L_x$ can be sharp
even when little or nothing is known about $L_y$ and $L_z$.
This statement of course holds true both in classical and in quantum
mechanics.
In this sense the $L_x$, $L_y$, and $L_z$ measurements are truly
independent measurements.

Between $V$ and $A$ measurement there appears to be a stronger
connection: it might be impossible to measure $A$ without knowing $V$.
This at least is the indication that emerges from a couple of examples
of area-measurement procedures.
Let us imagine that we measure the area of the surface
of a table using the time-of-flight of two light bursts
(I am assuming for simplicity that I have previously established that
the table is rectangular, so that by measuring two sides
I can obtain the area).
These would be two length measurements of the type
described in Subsection 3.3. The area should be obtained from two
time-of-flight measurements $T_1$ and $T_2$. But it is not sufficient
to measure $T_1$ and $T_2$ in order to obtain an area measurement:
it is necessary to know the velocity of the table!
If the table is at rest the area will be deduced from the $(T_1,T_2)$
measurement
as $A=T_1 {\cdot} T_2 {\cdot} c^2/4$. But if the area is moving with speed $V$
along the direction of the $T_1$ measurement procedure
one would instead deduce from the $(T_1,T_2)$ measurement
that\footnote{This assumes that the $T_1$ measurement is itself
independent of the knowledge of the speed of the table. In practice
it is most natural to set the clocks in rigid motion with the table,
and then the $T_1$ readout would have to be corrected
{\underline{in a $V$-dependent way}} by the observer, since the relevant
clock is not at rest with respect to the observer.
The additional $\sqrt{1-V^2/c^2}$ dependence does not change the nature
of the argument I am making, and I can therefore adopt the simplifying
assumption that $T_1$ is measured in a $V$-independent way.}
$A = T_1 {\cdot} T_2 {\cdot} (c^2 - V^2)/4$.

So, within this time-of-flight measurement procedure,
it is somewhat paradoxical to assume that one could get a sharp
area measurement without relying on a sharp surface-velocity
measurement. How could genuine area eigenstates not be simultaneously
surface-velocity eigenstates?
But if surface-area eigenstates are also surface-velocity eigenstates
then the discretization of areas would naturally be incompatible
with Lorentz symmetry.

In the case of $L_x$ measurement ({\it a la} Stern-Gerlach)
there is instead no paradox in assuming that $L_x$ would be found
experimentally to have discretized sharp values even when $L_y$ and
$L_z$ do not. There is no paradox in the ordinary-quantum-mechanics result
of ${\hat{L}}_x$ eigenstates which
are not ${\hat{L}}_y$ and ${\hat{L}}_z$ eigenstates.

The $V$ dependence may well be an artifact of the measurement procedure
here considered, and in fact I will not be able to argue for
this dependence in general. However,
as emphasized in Subsection~7.2, the need to know the velocity of the
surface is also present in the area-measurement procedure
based on a capacity measurement which had already been considered
in Ref.~\cite{carloarea}.
The same holds in
every area-measurement procedure
I considered: all of the ones I could think of
involved (more or less implictly, but in an inevitably
substantial way) a $V$ dependence of the map between the quantities
actually measured (times, length, capacities,...) and the area one
would attempt to evaluate through those measurements.
Interesting insight could be gained
if the community took the {\underline{challenge}}
of devicing an area measurement procedure truly free from
dependence on the velocity of the surface whose area is being measured.

In closing this subsection, since I have here introduced a second
area-measurement procedure, let me stress again some of the
points I already emphasized in Subsection~7.2, in the discussion
of the capacity-based measurement procedure.
Both area-measurement procedures satisfy my expectation that
the time of measurement required by area-measurement procedures
is at least of order $T \sim \sqrt{A}/c$. This I already discussed
for the capacity-based measurement procedure in Subsection~7.2.
and it is completely obvious in the area-measurement procedure
based on time-of-flight measurements.

Clearly also the area measurement procedure based
on time-of-flight measurements, just like the
area measurement procedure based
on capacity measurement,
relies on a distance measurement.
(And, as emphasized already, the status of distances in loop quantum
gravity is still being debated: it may well be impossible to
find an area operator with good eigenstates.)

Is worth emphasizing that
also the area measurement procedure based
on time-of-flight measurements, just like the
area measurement procedure based
on capacity measurement,
relies on the assumption that the surface be absolutely flat.
If the surface was only approximately flat the map
between the $(T_1,T_2)$ readout and the area of interest would only
be an approximate map, just like for the other measurement procedure
the relation between the capacity readouts and the area of the
surface is only approximate if the surface is not exactly flat.

Notice that some of the apparent obstructions to the observability
of the claimed area discretization that are encountered in
these measurement procedures have their root in the same
physical intuition which provided motivation for quantization
of geometry.
For example, at the level of formalism loop quantum gravity implements
the intuition that the spacetimes we perceive (with low-energy
probes) as classical are actually only approximately classical,
but then the fact that the metric is not classical affects
the maps $(T_1,T_2) \rightarrow A$ and $C \rightarrow A$
that are relevant for the two measurement procedures here considered.
And those maps are affected in a ``fuzzy" (uncontrolled) way:
the connection between the readout (respectively $(T_1,T_2)$ and $C$)
and the quantity we want to measure ($A$) is no longer sharp,
its validity is only approximate, so that an absolute limit
on the accuracy of the area measurement emerges.

\subsection{Fuzzyness and quantum-gravity measurement theory}
If indeed, as here conjectured, loop quantum gravity ends up
providing us the first theoretical framework for spacetime
fuzzyness, this result should probably be interpreted as
a result of the (partial) diffeomorphism invariance of the approach.
For decades there has been a portion of the quantum-gravity
community contemplating the possibility that quantum gravity,
as a result of the associated diffeomorphism invariance,
might require a new measurement theory.
These ideas have not captured the attention of the quantum-gravity
community as a whole, probably because the arguments used to support
them are often presented in a sloppy way.

A (unfortunately)
popular sloppy description of the reason why a new measurement theory
should be expected is based on two points: (a) the present measurement
theory requires an apparatus external to the system but ``nothing
is external to the gravitational field", and (b)
the present measurement theory requires decoupling between system
and apparatus but everything interacts with the gravitational field.
This is clearly an unsatisfactory line of argument. In fact
Rovelli in Ref.~\cite{carloarea} stressed that
the logical structure of measurement theory does not really
require an apparatus that is ``spatially external" to the gravitational
field, it just requires a separation between degrees of freedom which
are being studied (the system) and degrees of freedom that are used
to study them (the apparatus). So point (a) does not provide good
motivation for a new measurement theory.
Rovelli~\cite{carloarea} also stressed that
the present measurement theory does not require decoupling between system
and apparatus, on the contrary the apparatus must interact with the
system in order to be able to extract the sought information about the
system. In fact, measurements of electromagnetic fields
are done using charged probes. From this viewpoint
the fact that all probes are charged gravitationally
may be seen as convenient rather than armful for our present
measurement theory.
So also point (b) does not provide good
motivation for a new measurement theory.

The fact that most researchers advocating a new quantum-gravity measurement
theory resort to the weak (wrong) arguments (a)+(b)
has strongly penalized the understanding of this crucial point
by the quantum-gravity community as a whole. Sadly the correct and
strong argument in favour of a new quantum-gravity measurement
theory should be well known since some 70 years:
already in the mid 1930s Bronstein
realized~\cite{bro,stachelearly} that a key point of our present measurement
theory is the availability of the limit in which the charge that the probes
have with respect to the fields being measured has
vanishingly small effects as compared to the inertial
mass of the probes. In the appropriate sense (after appropriate dimensional
rescaling) a sharp measurement (such as the ones required to establish the
discreteness of the spectrum of a relevant observable)
is only possible in the limit
in which the ratio between the charge of the probe and the inertial
mass of the probe is vanishingly small. This has been studied in detail
for what concerns measurement of electromagnetic fields, where it
was established~\cite{rosen} that those fields can be sharply measured only
by using probes with vanishingly small ratio between
charge and inertial mass: $e/m_i \rightarrow 0$.
Brostein and Salomon (and several other authors in more recent times,
see {\it e.g.} Refs.~\cite{pap16,ngmpla})
realized that the Equivalence Principle renders
this limit unaccessible in the case of measurement of
gravitational fields, since the Equivalence Principle imposes
that the ratio between gravitational charge and inertial mass
cannot be varied at all: it is fixed to 1.
This obstruction that the  Equivalence Principle imposes to
our present measurement theory leads to the expectation that
geometic observables cannot be measured sharply, that there should be
an absolute limit to their measurability, that there should be
some fundamental fuzziness of geometric observables.
The considerations I made in this Section on the loop-quantum-gravity
area operator provide an explicit example of the difficulties
of sharp measurement in quantum gravity.

\subsection{On the Rovelli-Speziale operators}
In a study that progressed in parallel with the one I am here reporting,
Rovelli and Speziale have obtained some results~\cite{carlosimone}
that are relevant for some of the points I raised in this Section.
From my perspective,
Rovelli and Speziale report progress in realizing a first level
of analogy between the role of Lorentz symmetry in loop quantum gravity
and the role of space rotations in ordinary quantum mechanics.
For my arguments one concludes that such an analogy requires
noncommutativity between surface area and surface velocity
and this is indeed (although somewhat implicitly)
the type of structure that Rovelli and Speziale encountered.
They construct~\cite{carlosimone}
an operator $A'$ which could plausibly (although this is
a delicate point) describe
the boosted area as seen in the unboosted frame that defines
the canonical theory: if the canonical theory adopts the time
variable of observer $O$ it will also naturally host as an
observable the (equal-time) area $A$ of a surface, then this
same surface (more properly the same world-sheet)
will also define an equal-time area $A'$ for a boosted observer $O'$.
$O$ and $O'$ ``live" in different canonical theory (characterized
by different Hilbert spaces) since their spacetime foliation
(and therefore their time variable) are different, but the
comparison of the operators $A$ and $A'$ in different canonical
theories is not a natural concept.
Rovelli and Speziale propose to rewrite the operator $A'$
in terms of operators of the canonical theory of observer $O$.
At least in a certain limit (which in particular involves
infinitesimal boosts) they have a definite proposal
for the form of $A'$ in the canonical theory of $O$.
They then show that $A'$ does not commute with $A$
and they argue that this might render Lorentz symmetry
compatible with area discreteness. In fact, their
operator $A'$ when written in terms of observable of the
observer $O$ is basically (of course) a function
of $A$ and of the surface velocity $V$, therefore the fact that $A'$
does not commute with $A$ follows from the fact
that $V$ does not commute with $A$.

In Lorentz-symmetric {\underline{canonical}} theories
with area discreteness
the observation that $V$ should not commute with $A$ is perhaps more
fundamental than the observation that $A'$ should not commute
with $A$. In fact, the noncommutativity of $V$ and $A$ can be
formulated as a property of legitimate observables in the
same canonical theory, while the noncommutativity of $A'$ and $A$
must always rely on some alchemy that allows to compare
observables that live in different canonical theories.
Both $V$,$A$ noncommutativity and $A'$,$A$ noncommutativity
appear to allow $A$ discretization (after all, in a Lorentz-symmetric
theory, $A'$ should
roughly be an operator function of $V$ and $A$, although I provide
below some elements of caution on this point).
While my result on $V$,$A$ noncommutativity followed
exclusively from the analysis of the way in which Lorentz symmetry
governs the results of measurements\footnote{In fact my result
applies to any theory predicting discretized areas, whether or not
the theory makes use of the formalism of quantum mechanics.
What I mean by noncommutativity does not require the algebraic
concept of commutation relations between quantum-theory operators.
It simply corresponds to a statement about the results of
measurements: in a Lorentz-symmetric theory with area discretization
$V$ cannot be sharply measured when $A$ is sharply measured.},
the observation on $A'$,$A$ noncommutativity
reported in Ref.~\cite{carlosimone}
is a technical/algebraic argument drawn in the framework
of the formalism of quantum mechanics (applied to spacetime).
Therefore the discussion of $A'$,$A$ noncommutativity
proposed by Rovelli and Speziale follows
more closely the spirit of the already-mentioned
classic paper~\cite{snyder} by Snyder, in which the
compatibility of Lorentz symmetry with spacetime
discretization was first argued, with the important
difference that Snyder analyzed Lorentz-symmetry
transformations of distances (generously interpreting
Snyder's reference to coordinates), while Rovelli and Speziale
generalized that result to the case of areas.

Besides introducing a candidate description of the operator $A'$
in the canonical theory of observer $O$, Ref.~\cite{carlosimone}
also introduces some transformation operators $M$ that attempt to
reproduce Lorentz-symmetry transformations
connecting the operators $A$ and $A'$.
Under the assumption that $M$ is unitary
Rovelli and Speziale conclude that the spectrum of $A$ and $A'$
must be the same.

The operators $A'$, $M$ introduced by Rovelli and Speziale
represent an important development
for the loop-quantum-gravity approach, which indeed provides
encouragement for a first level of analogy
between the role of Lorentz symmetry in loop quantum gravity
and the role of space rotations in ordinary quantum mechanics.
Their analysis relies on some conjectures, especially concerning
the unitarity of $M$
and some implicit assumptions about a
surface-velocity operator.
The surface-velocity operator has not yet been rigorously
introduced in loop quantum gravity but
clearly governs the properties
of the $A'$ operator (from the perspective of observer $O$).
If these conjectures prove to be correct
the first level of analogy
between the role of Lorentz symmetry in loop quantum gravity
and the role of space rotations in ordinary quantum mechanics
will have been established and we will be left with the more delicate
issues which I also raised, together with $V$,$A$ noncommutativity,
in this section: those aspects of
Lorentz symmetry in a canonical theory which instead appear to
require that the analogy with space rotations in ordinary quantum
mechanics be limited.

An example of conjectured properties whose verification
deserves attention is found in the
interesting but delicate point of the study reported
in Ref.~\cite{carlosimone} which describes
the operator $A'$
that ``belongs" to observer $O'$ in terms of operators of
the $O$ observer. This description is obtained through a limiting
procedure (infinitesimal boosts...).
This attempt to describe the operators of $O'$ in terms
of operators of $O$ could be relevant for my concern about the
problem of the absence of the world-sheet from the formalism.
Lorentz symmetry naturally invites us
to make direct reference to the world-sheet and construct different
equal-time observables by suitable projections, but the canonical theory
would describe each of these projections as leaving in different
theories (different Hilbert spaces, different canonical theories
characterized by different time variable/parameter).
If we really succeed in giving a faithful description
of the operators of all observers in the terms of the operators
of one observer, perhaps the need of the world-sheet will be relaxed.
However, especially in light of its importance for the
physical problem here of interest,
this hypothesis must still be investigated in greater depth.
I fear that the canonical formalism will not allow us
to make genuine predictions concerning Lorentz symmetry:
rather than {\underline{showing}} that the measurements
done by $O'$ on the same world-sheet give the results
expected according to Lorentz symmetry (in comparison with the
results obtained by $O$ on that same world-sheet)
the canonical formalism will only allow us to {\underline{enforce
by hand}} a description of the operator $A'$ as a certain
function (roughly given by a function of $V$ and $A$)
of the operators that live in the canonical theory
of observer $O$. In this case we would
risk to missinterpret Lorentz symmetry as a prediction of
the theory, while instead we are just making use of our knowledge
of the rules of Lorentz-symmetry transformations to introduce
in the $O$'s canonical theory some observables (functions of
$V$ and $A$) which do not genuinely describe the Lorentz-transformed
observables (they might not truly describe observations performed by $O'$,
but rather describe some complicated functions of
natural observations, such as $V$ and $A$, conducted by $O'$).
In order to gain insight on these issues it may be useful
to push further the approach proposed by Rovelli and Speziale.
What happens if finite boosts are considered?
Is the description of $A'$ in terms of $O$'s operators
still satisfactory?
Can one set up a satisfactory Heisenberg (or Schroedinger)
picture if $O$ must make use of both ``his own operators"
and ``$O'$ operators"?
[For example, in the Schroedinger picture we want time-independent
observables and time-dependent states, but it might
be hard to describe $A'$ as time-independent
from the $O$ perspective, since according to $O$ the surface described
by $A'$ is not equal-time.]

As mentioned, I feel that,
together with these additional investigations of the
technical aspects of the Rovelli-Speziale approach,
the analysis reported in Ref.~\cite{carlosimone}
can provide a natural starting point for the investigation
of some of the delicate issues I raised in connection with
the limits that one must impose to the analogy
between the role of Lorentz symmetry in loop quantum gravity
and the role of space rotations in ordinary quantum mechanics.
In particular, Rovelli and Speziale
find encouragement for the idea that area eigenstates of
observer $O$ might be mapped into states that are not area
eigenstates for observer $O'$.
This renders even more urgent the point I raised concerning the
fact that Lorentz symmetry (unlike space-rotation symmetry)
also governs some measurements
(see Subsections~3.3, 3.4, 3.5)
in which two observers share the measurement procedure: both observers
use the same readout (although they adopt different calibrations)
to measure the same area. How can then two observers
disagree on whether the results can be characterized
with an area eigenstate?

\subsection{Field theory and covariant reformulations of loop quantum
gravity}
Some of the concerns I discussed here and some of the points that must
still be addressed, even after the important developments in
the loop-quantum-gravity
formalism reported by Rovelli and Speziale~\cite{carlosimone},
have to do with the interplay between space/time transformations,
like Lorentz boosts,
and the structure of a canonical quantum theory, with the peculiar
role attributed to the time variable/coordinate/parameter.

It is difficult to formulate a conjecture
on the outlook of these issues
in a covariant reformulations of loop quantum
gravity, such as the ones being already explored relying on the
so-called ``sum over histories" formalism~\cite{carloliving}.
Perhaps this covariant reformulation will lead to a satisfactory
description of equal-time observables in terms of world-sheets,
world-lines and their relations (intersections...).

Some insight might be gained by
using as guidance the
success of field theory in background Minkowski spacetime,
which is indeed a quantum theory with Lorentz symmetry.
The guidance provided by field theory will however be
limited, since field theory lives in classical
and nondynamical spacetime and of course
it makes no quantum predictions
concerning spacetime observables (such as areas, volumes...).

\subsection{Specifically on discretization of rest area}
In the preceding subsections I have presented my case for
area fuzzyness in loop quantum gravity.
For completeness let me return to the hypothesis (which however
I disfavour) that the area discretization is observable.
And let me focus on the special case in which the spectrum
discussed in the loop-quantum-gravity literature would only apply
to the case of an area at rest.

As clarified above, Lorentz symmetry should be abandoned if that
same spectrum applied to areas of all velocities
(sharply determined velocities).
It is instead possible to assume that the presently-adopted
spectrum applies only to areas at rest
(sharply determined to have zero velocity). However, it would then
be necessary to assume/predict that boosted observers
(observers that assign a nonvanishing velocity to the area)
would see a boosted picture of that spectrum
(including a Lorentz-Fitzgerald contraction
of the discretization scale).
The area operator and its spectrum
should depend on the velocity of the area.
One might be tempted to think otherwise, for example assuming
that Lorentz symmetry might be preserved if the same physical
surface could be an area eigenstate for one observer and
not an area eigenstate for another observer (this would in principle
allow that the spectrum be the same for all observers, even the ones
in motion with respect to the area); however,
consistency with classical Lorentz symmetry imposes
that an area eigenstate with well-specified velocity of the
surface (in particular, ``at rest", $V=0$) {\underline{must}}
be an area eigenstate
with well-specified velocity for all other inertial observers.

I find that the idea the the area operator which is being adopted by
the loop-quantum-gravity research programme would refer
to the rest area of a surface is peculiar also from a technical
perspective: the structure of that operator only makes reference
to some space coordinates that delimit the area. It includes no
reference to the fields that physically identify that surface
and it makes no reference to the time evolution of the surface.
As required by canonical quatization, it is just an operator that
refers to quantities defined at a fixed time. The operator does not
specify in any way where the surface will be at a later
time. At a later time (however small is the time difference)
the surface could well be still identified by those coordinates,
but it might equally well have moved somewhere else, and in both cases
the area operator would just take the fixed-time section of the
world-sheet identified by the surface and calculate its area
according to a certain prescription.

\subsection{On Lorentz-symmetry deformations and on the Gambini-Pullin
mechanism for induced violations of Lorentz symmetry}
In closing this Section of the fate of Lorentz symmetry in loop
quantum gravity I find appropriate to emphasize the differences
between the type of departure from Lorentz symmetry I considered
here and certain other studies of departures from Lorentz
symmetry.
I have been concerned with the status of Lorentz symmetry
at the fundamental theory level.
In a series of papers initiated by a study
by Gambini and Pullin~\cite{gampul,mexpap}
deviations from Lorentz symmetry motivated by loop quantum gravity
had already been considered, but these are of totally different nature.
Gambini and Pullin rely on the introduction of a background of
a ``weave state"~\cite{weave}. The type of Lorentz-symmetry violation
they discuss is the very familiar one encountered whenever a
background is introduced. It is a departure from Lorentz symmetry
which is {\underline{induced}} by the presence of the background,
rather than being truly fundamental in the theory.
The Gambini-Pullin mechanism is the one we are familiar
with from the study of the
behaviour of light in water, crystals, and other media and also from
theory work on certain fixed-background spacetime
(such as the canonical noncommutative spacetimes discussed
in the preceding Section).
The Gambini-Pullin mechanism in fact is not in any way related
with the discretization of areas/volumes.
On the contrary my considerations on the fate of Lorentz
symmetry in loop quantum gravity
concern the truly fundamental level of analysis of the theory.
I am not considering any special choice of background.
I am looking for some fundamental
implications of the spacetime discretization
encountered in the loop-quantum-gravity formalism.

The nature of my considerations on the fate of Lorentz
symmetry in loop quantum gravity is also not directly connected with
the proposal I put forward~\cite{dsr1dsr2} of deformations of
Lorentz symmetry.
While, as discussed in Subsection~6.1, those deformations should
provide the exact/fundamental description of the symmetries
of certain noncommutative spacetimes, I expect that
their applicability to
loop quantum gravity, if any, should be confined to contexts in
which some level of coarse-graining of the fundamental spacetime
picture has been advocated.
It appears to be plausible that
a deformation of Lorentz symmetry might emerge in the study of
the limiting procedure that the loop-quantum-gravity
research programme must identify
in order to make contact with classical spacetimes.
In fact, while the classical limit must be described by ordinary
classical symmetries (in particular classical flat spacetimes
should have Lorentz invariance),
if one stops the limiting procedure a bit before the classical limit,
where spacetime would already look ``quasi-classical",
it is plausible that a deformation of Lorentz symmetry
would play a role.
In this author's opinion the classical-spacetime limit should
correspond to the limit in which the particles that probe
spacetime have extremely-low energy (extremely-large wavelengths).
For probes of finite but low energy (``moderately-large" wavelength)
spacetime would be perceived as ``quasi-classical" (and possibly
characterized by a deformation of Lorentz symmetry),
but in the low-energy limit
a truly classical spacetime would emerge (together with its ordinary
classical symmetries, including Lorentz invariance of flat spacetimes).

\section{Closing remarks}
It is not accidental that in this study the number of issues
that have been raised is much larger than
the number of issues for which an answer has been provided.
While it is not difficult to analyze
the fate of classical spacetime symmetries in
theories in which nonclassical properties are only assigned to
non-spacetime degrees of freedom,
the analysis of spacetime symmetries in nonclassical pictures of
spacetime itself involves a large number of subtle issues.
In some cases there is not even an {\it a priori}
intuitive way to rephrase the questions we usually ask of
a spacetime symmetry in classical spacetime. Perhaps the best example
of this is the concept of spontaneous breaking of spacetime symmetries:
we are familiar with spontaneous symmetry breaking in particle-physics
theories, where the concept of ``vacuum" is clear to us,
but we lack any depth in the understanding of the ``spacetime vacuum".

In the type of noncommutative geometries here considered,
using some recent mathematical-physics results, it is now
clear that classical Lorentz symmetry does not hold.
For canonical spacetimes, the simplest case from the technical
perspective, we even have a rather deep understanding of the
relevant type of violation of Lorentz symmetry, but, unless
we are willing to accept the existence of truly preferred observers,
we must devote urgent attention to the search of
spontaneous-symmetry-breaking mechanisms that might support it.
In $\kappa$-Minkowski Lie-algebra spacetime it is clear that
the appropriate concept of transformation rules between
inertial observers will require the concept of deformed Lorentz
symmetry (a group of finite transformations)
introduced in Refs.~\cite{dsr1dsr2},
and it is clear that (at least in the one-particle sector~\cite{dsr1dsr2})
infinitesimal symmetry transformations
should be described by a Hopf algebra of the type considered
in Refs.~\cite{majrue,kpoinap}. But the precise description
of the symmetries of $\kappa$-Minkowski still requires further study;
in particular, certain mathematical-physics
studies~\cite{kpoinap,jurekdsrNEW,lukiedsr} appear to argue
that several Hopf algebras are equally good candidates
as tools for the description of the symmetries of $\kappa$-Minkowski
spacetime. This is clearly unacceptable physically:
we are allowed to introduced meaninfully
the concept of ``symmetry of a spacetime"
only if we are able to associate to a given spacetime one and only
one mathematical structure which describes its symmetries.
This is another issue that deserves urgent investigation,
paricularly since preliminary estimates~\cite{polonpap}
of the departures from
ordinary Lorentz symmetry required by $\kappa$-Minkowski suggest
that forthcoming experiments~\cite{glast} should be able
to test the prediction.

The case of loop quantum gravity is even more interesting.
While the examples of noncommutative spacetimes I considered
have relatively narrow objectives (at best providing an
effective description of spacetime in an appropriate quasi-classical
and quasi-flat limit) the loop-quantum-gravity research programme
is constructing a full ambitious description of quantum gravity.
As the issue of Lorentz symmetry gains importance in light
of the developing experimental situation, the loop-quantum-gravity
research programme discovers to be unprepared to provide this
key prediction. As shown here the loop-quantum-gravity
analysis of the concept of area in quantum spacetime
is insufficient for providing guidance on this important
issue, in spite of the emergence of discretization.
It is on this issue of the description of areas in loop
quantum gravity and the associated analysis of Lorentz
transformations that new technical and conceptual
developments appear to be most urgently needed.

Concerning loop quantum
gravity my analysis has provided both material in
support of preservation of ordinary Lorentz symmetry
and material in support of loss of Lorentz symmetry.
My point that area discretization is not in general
inconsistent with Lorentz symmetry might be of encouragement
for the idea that Lorentz symmetry be preserved in loop
quantum gravity, but it opens the challenge of a proper
definition of the observable ``velocity of the surface".
While at first sight the construction of this observable
does not appear to be confronted with in-principle obstructions,
some of the observations I reported here about the interplay between
surface velocity and surface area in the way in which Lorentz
symmetry is realized in our (present) observations may suggest
that some obstacle for Lorentz symmetry might be eventually
encountered in loop quantum gravity.
I have analyzed various sources of concern for
the compatibility of Lorentz symmetry
with the type of area discretization
predicted by loop quantum gravity, which indeed are all in some way
connected with the interplay between surface velocity and surface
area, and from a physical/operative perspective originate from
the following two challenges:
(i) the need to identify a area-measurement procedure which is not affected
by the surface-velocity uncertainty (otherwise the noncommutativity
between surface velocity and surface area, needed for
the compatibility between
area discretization and Lorentz symmetry, will end up
rendering unobservable the discreteness of the area spectrum), and
(ii) the need to understand the
role that measurement procedures such as the ones described
in Subsections~3.3, 3.4, 3.5, in which the readouts
of the same measurement procedure are meaningful for different
inertial observers, should have in a theory in which
area eigenstates for one inertial observer are predicted to
be mapped into superpositions of area eigenstates for
another inertial observer. (On the one hand eigenstates for $O$
must go into superpositions of eigenstates of $O'$ if area
is discretized but the expectation values of the area observable
satisfy the usual FitzGerald-Lorentz contraction, on the other
hand measurement procedures such as the ones described
in Subsections~3.3, 3.4, 3.5 appear to require that eigenstates
of length/area/volume should be eigenstates for all
inertial observers.)

The construction and careful analysis of
the observable ``surface velocity" in loop quantum gravity
is the key to the understanding
of the fate of Lorentz symmetry in that theory and is also
important for establishing whether or not area discreteness is,
at least in principle, observable.
Through the understanding of the properties of this, still unknown,
surface-velocity observable we could establish whether
its commutation relations with the surface-area observable
are of a type that may render area discreteness compatible
with Lorentz symmetry, and, perhaps, we could also gain
some insight on the issues related with
my concerns for the operative definition
of the area observable.

Very little of what I observed in
Section~7 is specific to loop quantum gravity: my observations
apply to any canonical quantum theory of gravity
with area discretization.
My observations all originate from the fact that,
in the example of the area observable, the primary/fundamental
entity from the Lorentz-symmetry perspective is the surface's
world-sheet, but in a canonical quantum theory one can only
meaningfully introduce observables that are defined at a fixed time.
The programme of investigation of loop quantum gravity
for which the present study provides motivation
might also provide insight on the wider
subject of the interplay between area/volume discretization
and Lorentz symmetry in other canonical quantum theories
of gravity.

\section*{Acknowledgments}
As this written notes were being prepared I benefitted
from conversations with several colleagues.
Special thanks I owe to
Dario Benedetti,
Steve Carlip,
Jurek Kowalski-Glikman,
Fotini Markopoulou,
Jack Ng,
Michael Reisenberger,
Carlo Rovelli,
Lee Smolin,
Simone Speziale
and
Hendrik van Dam.
I am particularly greatful to Jurek for discussions
on some aspects of Snyder's work~\cite{snyder} and to
Michael for discussions on various aspects of Lorentz symmetry.
The availability of Dario and Simone to le me ``try on them" some of
the arguments here presented was especially important for my
work. In particular, conversations with Dario helped me become more
confident that the classical language adopted in Subsection~2.4
does have some (however limited) usefulness, while conversations
with Simone made me realize that among my arguments in support
of ``fuzziness" (against ``observable discretization")
one of the most effective (from the communication viewpoint)
is the one that observes that the same
physical intuition that motivates
the description of area as a quantum operator should also motivate
a reanalysis of the operative definition of area.

\baselineskip 12pt plus .5pt minus .5pt

\vfil

\end{document}